# Non-resonant SABRE provides a new and versatile hyperpolarization approach in magnetic resonance

[1]Loren L. Smith, [2]Warren S. Warren*
[1]Department of Chemistry, Duke University, Durham, NC 27712
[2]Departments of Chemistry, Physics, Biomedical Engineering, and Radiology, Duke University, Durham, NC 27712

Abstract: Hyperpolarization approaches in magnetic resonance overcome the sensitivity limitations imposed by thermal magnetization and play an important role in a very wide range of modern applications. One of the newer strategies, variants of what is generically called SABRE, uses para-hydrogen to form hydrides on transition metal catalysts, followed by reversible exchange to polarize target molecules in solution, and has produced large signal enhancements ($\approx 10^4$) on hundreds of different molecules, cheaply and rapidly. Most commonly, the sample is kept in a constant field, matched to make the hydride scalar coupling comparable to the frequency difference between hydride protons and target protons ($\approx 6.5$ mT) or hydride protons and target heteronuclei ($\approx 0.5$ $\mu$T). Here we demonstrate a different strategy, applicable to a wide range of target molecules, that produces field-independent spin order in the target molecules that is efficiently converted to magnetization. The observed signal is even independent of field direction, hence significant polarization can be achieved in a sample on a lab bench with no field control at all. We show this signal arises from creation of two-spin order in the target molecules, and discuss multiple ways this strategy should expand SABRE generality and efficiency. We also show that, in many cases, the standard assumption in low-field SABRE of a starting state with only singlet polarization leads to incorrect results.

Introduction

Sensitivity is almost always a limitation in magnetic resonance, because the small energy difference between nuclear spin states (gyromagnetic ratio γ times field $B$) limits thermal magnetization. For $^1$H at 15 Tesla, $\gamma B/k_B T \approx 10^{-4}$; for isotopes with lower γ ($^{13}$C, $^{15}$N) it is much worse. "Hyperpolarization" (HP) methods solve this problem by dramatically increasing the fractional magnetization (and hence the sensitivity), typically by $10^3$-$10^5$. This paper focuses on one of the newer methods, variants of SABRE (signal amplification by reversible exchange) and identifies a significant omission in the general understanding of this method: it neglects creation of two-spin order in target molecules. Correcting this emission should lead to signal enhancements and new experimental strategies.

The obvious hyperpolarization method, drastic cooling (mK-K) in a strong magnetic field[1] is extremely slow[2]. Thus, almost all hyperpolarization technologies get the nuclear spin order from something else, most commonly unpaired electron spins. Examples include optical pumping (nitrogen vacancy centers in diamond,[3] electron spins in Rb vapor followed by collisions to polarize noble gases [4-9], or triplet states in organic molecules)[10-15]. Another strategy is to introduce a paramagnetic species (generally called dynamic nuclear polarization (DNP)),[16-22] sometimes by ionizing[23] or UV radiation[24], more often by doping the target of interest with a radical[25]. In the cryogenic case, the sample is then rapidly dissolved in warm solution (dissolution DNP or d-DNP)[21, 26-34]. The first clinical DNP application[35], metabolism of $^{13}$C-labeled pyruvate to alanine or lactate, has been validated as a prostate cancer marker and is currently in clinical trials, and other molecules such as $HCO_3^-$ also show promise[36-38]. While all of these methods are being actively improved, none of them approach the level of simplicity, generality and convenience needed to enable (for example) hyperpolarization as a routine improvement in NMR or MRI.

Parahydrogen ($p$-H$_2$, the singlet state $S \equiv (|\alpha\beta\rangle - |\beta\alpha\rangle)/\sqrt{2}$ of gaseous H$_2$) provides a different source of spin order. It is an extraordinary "quantum reagent" in a metastable, pure spin state which can be made in large quantities at high spin purity and low cost[39-44] and stored stably for weeks.[45] $p$-$H_2$ was initially mostly used in NMR by addition across inequivalent carbon double or triple bonds, then pulse sequences converted singlet order into nuclear polarization[46-51], but each target molecule requires different optimization (often including a different catalyst) and there is a very limited range of targets. In 2009, Duckett and coworkers introduced SABRE (Fig. 1)[52-56] which works by reversibly binding $p$-$H_2$ and target ligands (such as pyridine) to an octahedral transition metal complex, usually iridium-based. Under the right circumstances (see Fig. 1), para order in the hydrides is converted to positive ligand (and negative hydride) magnetization. After both $p$-$H_2$ and ligand exchange (typically the ligand exchange is much faster), the Ir atom can catalyze another transfer. SABRE directly polarizes hydrogen on the ligand target; methods for direct heteronuclear polarization such as SABRE-SHEATH[57] go by the general name X-SABRE (there are now at least a dozen such methods with a wide range of applications).

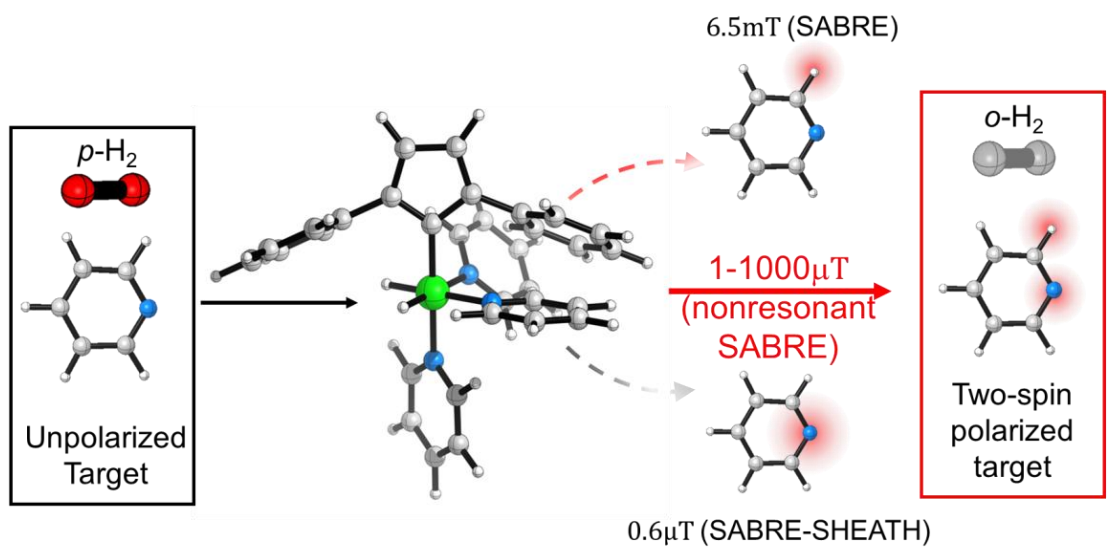

Fig. 1. Comparison of different SABRE variants. In all cases $p$-H$_2$ and a target ligand (here pyridine) bind reversibly in the equatorial positions of an Ir complex; typically, one axial position has a large ligand such as iMes illustrated here. In SABRE, four-bond couplings between the hydride and target $^1$H atoms create nuclear spin polarization, commonly 1000-10,000 times larger than what is observed at thermal equilibrium in even a large magnet (only one proton is highlighted for simplicity). In X-SABRE methods, the much larger two-bond couplings between the hydride and $^{15}$N create polarization. In this paper we demonstrate a third mechanism, nonresonant SABRE, with direct creation of heteronuclear two-spin order in the ligands. This has vastly different properties from the first two approaches.

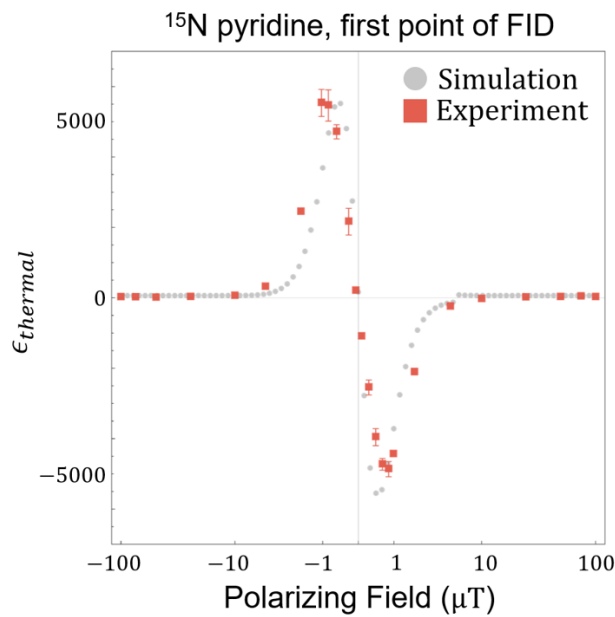

Fig. 2. Signal enhancement $\epsilon$ for $^{15}$N-pyridine in Ir(H)$_2$(IMes)(pyr)$_3$ with continuous application of static polarizing field $B$. Gray points show theoretical simulation of a Ir(H)$_2$(IMes)(pyr)$_3$ system ($k_L$ = 16 s$^{-1}$, $k_H$ = 2 s$^{-1}$, [catalyst]: [ligand] = 1 : 20, and 5-s evolution). Red data points show experimental magnetization enhancement relative to thermal magnetization measured at 1 T (measured as the amplitude of the FID immediately after a 90$^\circ$ pulse).

SABRE and X-SABRE avoid all cooling, can hyperpolarize in seconds, are not as limited in volume as DNP, and have been applied to a wide variety of molecules. Today many groups work on different aspects including methods that use shaped three-dimensional magnetic fields, or work in a high field magnet using rf pulses[58-77]. However, the "workhorse experiment" since the beginning has used a constant magnetic field and no rf irradiation. There are two basic regimes: the traditional SABRE case, where the field is matched to make the hydride scalar coupling comparable to the frequency difference between hydride protons and target protons (≈6.5 mT), and the SABRE-SHEATH case, where the field is matched to the frequency difference between hydride protons and target heteronuclei such as $^{15}$N and $^{13}$C (≈0.5 µT, achieved in a mu-metal shield[57]). In both cases, the matching condition is generally justified as an example of level anticrossings, although this does not quite work in the limit of large couplings as has been detailed elsewhere[78]. Still, as Fig. 2 shows, well-validated simulations[78, 79] reproduce the experimental field dependence well. Constant-field experiments are constrained by inconvenient field strengths that require a mu-metal shield to achieve, and the high field experiments are very sensitive to experimental complications such as resonance frequency difference between the two hydride spins.

Here we show (to be honest, to our surprise) that this picture is an oversimplification. For a wide range of target ligands which have both protons and heteronuclei (including the most heavily studied molecules in the field), there is a highly efficient path to *two-spin* order in the target which can be quantitatively converted to observable magnetization competitive with SABRE and X-SABRE. Over the field regime between SABRE and X-SABRE, this produces magnetization independent of field strength and direction — in other words, inverting the field has no effect on the magnetization. For this reason, we call this effect nonresonant SABRE. Magnetization can even be produced by bubbling directly on a tabletop with no external magnet or shield. To our knowledge this mechanism has never been considered, and it also opens up many options for SABRE improvement in other regimes.

Results

We start by presenting experimental evidence for nonresonant SABRE. Figure 3 shows hyperpolarized FIDs and spectra for 160 mM $^{15}$N-pyridine at two specific fields, one which matches the well-known SABRE-SHEATH condition and one that is about a factor of 100 away. As the field strength increases, the spectrum transitions smoothly from in-phase to antiphase, then remains constant over a wide variety of field strengths. Far from resonance, indeed the FID is zero immediately after a 90 pulse (implying there is no one-spin z magnetization), but the FID grows in at later times. The first maximum occurs at $t = 1/4J_{NH}$ where $J_{NH}$ is the coupling between the pyridine $^{15}$N and the ortho protons ($\approx -10$ Hz).

Fig. 4 shows this field dependence over about three orders of magnitude. As shown in the SI, detection of $^1$H magnetization instead of $^{15}$N magnetization reveals signal comparable to the $^{15}$N magnetization off resonance, with similar antiphase behavior.

We note several experimental observations which contribute to the analysis below.

1. We stopped the bubbling for a variable period prior to pulsing the sample after moving it from the polarizing field to the 1 T measurement magnet, and saw much faster attenuation of the off-resonance signal ($T_1 = 12.46 \pm 0.26$ s after polarization build-up at 50 µT) than was observed for the conventional SABRE-SHEATH signal ($T_1 = 96.9 \pm 5.9$ s after polarization at 0.5 µT).

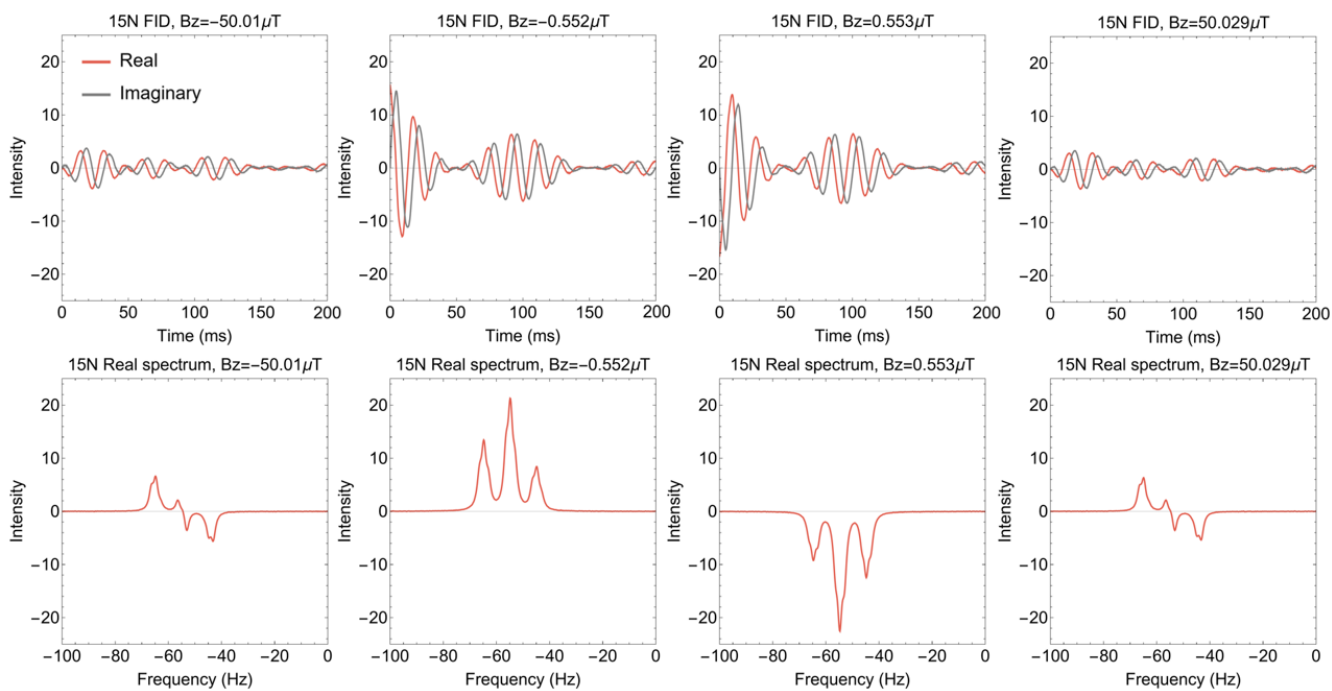

Fig. 3. Experimental one-scan $^{15}$N signal from 160 mM $^{15}$N-pyridine after excitation at the peak of the SABRE-SHEATH resonance (0.5 µT), and a factor of 100 away from that resonance (50 µT). The off-resonance signal is nearly zero immediately after the pulse, implying that no z-magnetization was directly produced. However, substantial signal grows in at later times. The amplitude and phase of the FID of the off-resonance signal is nearly independent of field over the range 5-500 µT, including inverting the field. $^1$H or $^{15}$N π/2 pulses applied after the polarization time produce nearly equal $^1$H or $^{15}$N fractional magnetization (respectively, see SI).

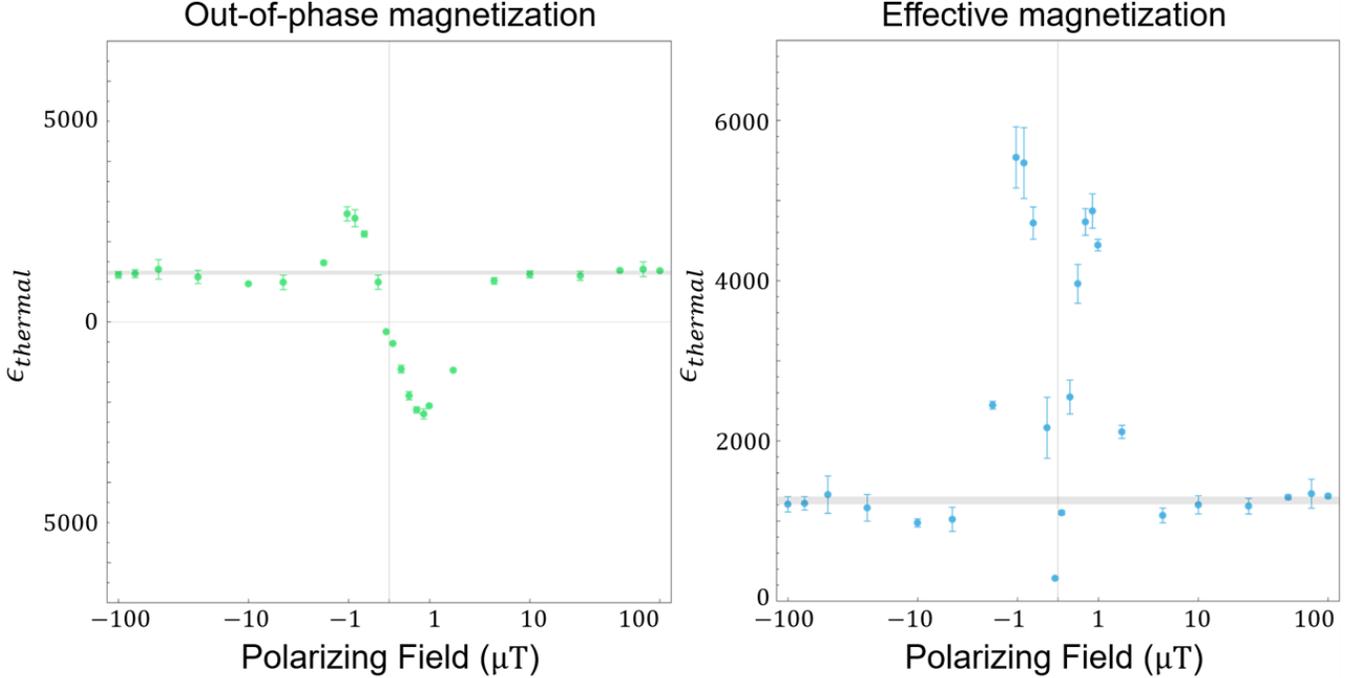

Fig. 4. Experimental $^{15}$N SABRE signals as a function of polarizing field. At left, the signal at a time $t = 1/4J_{NH}$ is plotted; this time corresponds to the maximum signal in the off-resonance case. The right plot shows the "effective magnetization", corresponding to the maximum value of the FID (which would also be the largest z magnetization available after an additional π/2 pulse). This is mathematically equivalent to acquiring the spectrum, optimizing the zero-and first-order phase, and taking its integral. The gray bars indicate the 95% confidence interval for the mean out-of-phase signal at high ($> 25 \mu T$) fields. Enhancement ε was calculated based on the maximum intensity of a thermally polarized reference sample (see SI).

2. We performed SABRE at three different points in the lab, without a shield, all with roughly 80 μT field aligned significantly away from the z-axis of the magnet used to detect the magnetization, and all gave comparable antiphase signal.

3. As seen in Figure 4, inverting the direction of the field has no effect on the off-resonance FID.

Theoretical basis for nonresonant SABRE

In this section, we will focus on the simplest spin topology that reveals the nonresonant SABRE effect. This shows that the experiments above provide compelling evidence of a hitherto unappreciated SABRE mechanism which directly and efficiently produces two-spin z order in the target ligand.

We start by summarizing the theoretical approach of ref. [78] for SABRE-SHEATH, as the modifications needed to understand nonresonant SABRE follow directly (Fig. 5). SABRE-SHEATH and SABRE differ only in the assignment of the $L$ nucleus (on the attached ligand) as $^1$H or a heteronucleus, typically $^{15}$N or $^{13}$C. The initial density matrix is assumed to have singlet order between the two hydride protons, derived from binding para-hydrogen from solution, but otherwise is random.

$$\hat{\rho}_0 = \left(\frac{1}{4}\hat{1} - \vec{I}_1 \cdot \vec{I}_2\right) \otimes \hat{1} \quad [1]$$

The simplest SABRE case is a three-spin system, where the spin operator of the target nucleus is labelled as $\vec{L}$ (which is $^1H$ for SABRE, and a heteronucleus for SABRE-SHEATH) and the two hydride spins are $\vec{I}_1$ and $\vec{I}_2$. As shown in ref. [78], the three-spin system (which is equivalent to assuming only one of the equatorial positions is occupied by an NMR-active species) gives similar results at this level to the four-spin system (which assumes both positions are occupied) so for simplicity we will stay with this model here.

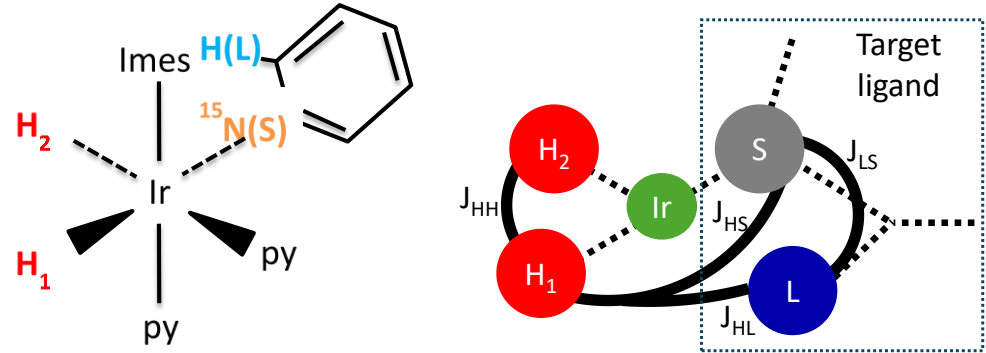

Figure 5. Simplified schematic to explain both resonant and non-resonant SABRE. In the simplest case, para-H2 and one NMR-active ligand are bound in the equatorial plane of the octahedral Ir complex. The Ir is commonly bound to a lone pair (e.g. $^{15}N$ or a carbonyl group). In both SABRE and SABRE-SHEATH, this model includes one ligand L that is scalar coupled to hydrogen $H_1$. Because of the geometry, only the linear H-Ir-ligand geometry supports a scalar coupling. For nonresonant SABRE, we assume L is $^1H$ and S is $^{15}N$ or $^{13}C$. The left side illustrates the topology for pyridine, one of the molecules studied extensively in this paper; however, all that matters in our analysis is the coupling network, shown more explicitly at right.

The nuclear spin Hamiltonian has the form:

$$\hat{\mathcal{H}}(t) = -(B_0)\left(\gamma_H(\hat{I}_{1z} + \hat{I}_{2z}) + \gamma_L \hat{L}_z\right) + 2\pi J_{HH}(\vec{I}_1 \cdot \vec{I}_2) + 2\pi J_{HL}(\vec{I}_1 \cdot \vec{L}) \quad [2]$$

here in natural units ($\hbar = 1$). For simplicity, we couple the ligand spin to only one of the hydride spins (spin 1), as this is consistent with experimental observations that only the hydride directly opposite the ligand has an observable coupling. We can rearrange this Hamiltonian as

$$\hat{\mathcal{H}}(t) = -(B_0)\gamma_H(I_{1z} + I_{2z} + L_z) + (\Delta\omega_{HL})L_z + 2\pi J_{HH}(\vec{I}_1 \cdot \vec{I}_2) + 2\pi J_{HL}(\vec{I}_1 \cdot \vec{L}) \quad [3]$$

in which $\Delta\omega_{HL} = B_0(\gamma_H - \gamma_L)$. The first term commutes with the rest of the Hamiltonian (and with the operators we are going to prepare) so we drop it in what follows for simplicity; this is equivalent to viewing the system in the $^1H$ hydride rotating frame, but without explicit assumptions on the field strength. We will thus use the reduced Hamiltonian

$$\hat{\mathcal{H}}' = (\Delta\omega_{HL})L_z + 2\pi J_{HH}(\vec{I}_1 \cdot \vec{I}_2) + 2\pi J_{HL}(\vec{I}_1 \cdot \vec{L}) \quad [4]$$

The coherence transfer pathway leading from singlet order on the hydrides to magnetization on the target nucleus goes through 4 commutators starting with the initial density matrix in equation [1]. Thus a standard Taylor series expansion for the time evolution of the density matrix at $t = 0$ ($d\hat{\rho}/dt = i[\hat{\rho}_o, \hat{\mathcal{H}}]$; $d\hat{\rho}^n/dt^n = i[d\hat{\rho}^{n-1}/dt^{n-1}, \hat{\mathcal{H}}]$) shows that the first term in the density matrix which has ligand magnetization $L_z$ appears in the fourth derivative. Pathways which produce this term originate from $I_{1z}I_{2z}$ (eq. [5-8]) or from $I_{1x}I_{2x}$ and $I_{1y}I_{2y}$ (eq. [9-12]).

$$I_{1z}I_{2z} \xrightarrow{J_{HL}(I_{1y}L_y)} -iI_{1x}I_{2z}L_y \xrightarrow{J_{HH}(I_{1z}I_{2z})} -I_{1y}L_y \xrightarrow{\Delta\omega_{HL}L_z} -iI_{1y}L_x \xrightarrow{J_{HL}(I_{1y}L_y)} L_z \quad [5]$$

$$I_{1z}I_{2z} \xrightarrow{J_{HL}(I_{1y}L_y)} -iI_{1x}I_{2z}L_y \xrightarrow{\Delta\omega_{HL}L_z} I_{1x}I_{2z}L_x \xrightarrow{J_{HH}(I_{1z}I_{2z})} -iI_{1y}L_x \xrightarrow{J_{HL}(I_{1y}L_y)} L_z \quad [6]$$

$$I_{1z}I_{2z} \xrightarrow{J_{HL}(I_{1x}L_x)} iI_{1y}I_{2z}L_x \xrightarrow{J_{HH}(I_{1z}I_{2z})} -I_{1x}L_x \xrightarrow{\Delta\omega_{HL}L_z} iI_{1x}L_y \xrightarrow{J_{HL}(I_{1x}L_x)} L_z \quad [7]$$

$$I_{1z}I_{2z} \xrightarrow{J_{HL}(I_{1x}L_x)} iI_{1y}I_{2z}L_x \xrightarrow{\Delta\omega_{HL}L_z} I_{1y}I_{2z}L_y \xrightarrow{J_{HH}(I_{1z}I_{2z})} iI_{1x}L_y \xrightarrow{J_{HL}(I_{1x}L_x)} L_z \quad [8]$$

$$I_{1x}I_{2x} \xrightarrow{J_{HL}(I_{1y}L_y)} iI_{1z}I_{2x}L_y \xrightarrow{J_{HH}(I_{1x}I_{2x})} -I_{1y}L_y \xrightarrow{\Delta\omega_{HL}L_z} -iI_{1y}L_x \xrightarrow{J_{HL}(I_{1y}L_y)} L_z \quad [9]$$

$$I_{1x}I_{2x} \xrightarrow{J_{HL}(I_{1y}L_y)} iI_{1z}I_{2x}L_y \xrightarrow{\Delta\omega_{HL}L_z} -I_{1z}I_{2x}L_x \xrightarrow{J_{HH}(I_{1x}I_{2x})} -iI_{1y}L_x \xrightarrow{J_{HL}(I_{1y}L_y)} L_z \quad [10]$$

$$I_{1y}I_{2y} \xrightarrow{J_{HL}(I_{1x}L_x)} -iI_{1z}I_{2y}L_x \xrightarrow{J_{HH}(I_{1y}I_{2y})} -I_{1x}L_x \xrightarrow{\Delta\omega_{HL}L_z} iI_{1x}L_y \xrightarrow{J_{HL}(I_{1x}L_x)} L_z \quad [11]$$

$$I_{1y}I_{2y} \xrightarrow{J_{HL}(I_{1x}L_x)} -iI_{1z}I_{2y}L_x \xrightarrow{\Delta\omega_{HL}L_z} -I_{1z}I_{2y}L_y \xrightarrow{J_{HH}(I_{1y}I_{2y})} iI_{1x}L_y \xrightarrow{J_{HL}(I_{1x}L_x)} L_z \quad [12]$$

These pathways result in direct magnetization on the target nucleus $\hat{L}_z$ with a $t^4$ dependence[78]:

$$\dddot{\rho}\frac{t^4}{4!} = 8\pi^3 J_{HL}^2 \Delta\omega_{HL} J_{HH} \frac{t^4}{4!}(2\hat{L}_z) + \cdots \quad [13]$$

This approach shows why a nonzero external magnetic field is necessary. It comes into this derivation in the $\Delta\omega_{HL}$ term by generating a phase shift which is needed to create the final magnetization. Mathematically, this happens because it creates a resonance frequency difference between the hydride and the ligand. Note that in every case in equations [5-12], the "flip-flop" component *(xx+yy)* of the coupling $J_{HL}$ appears as the first and fourth commutator, with the $\Delta\omega_{HL}$ and $J_{HH}$ terms appearing in between (in either order, as they commute with each other).

This approach does not show the optimal field, which was originally found by invoking level anti-crossing arguments. As ref. [78] shows, this prediction generally works when the $J_{HH}$ coupling dominates (as in SABRE) but can be off by as much as an order of magnitude in the SABRE-SHEATH case. However, well-validated numerical calculations starting from the equilibrium density matrix agree with equation [13], and agree with experimental data (Fig. 2).

We now turn to the nonresonant SABRE case. The nucleus *L* is now assumed to be a proton in the ligand, and we now add one more nucleus *S* assumed to be a heteronucleus on the ligand. This adds the terms in the bracket below.

$$\hat{\mathcal{H}}' = 2\pi J_{HH}\vec{I}_1 \cdot \vec{I}_2 - \Delta\omega_{HL}L_z + 2\pi J_{HL}(\vec{I}_1 \cdot \vec{L})$$
$$+\{2\pi J_{HS}I_{1z}S_z + 2\pi J_{LS}(S_xL_x + S_yL_y) + 2\pi J_{LS}S_zL_z + 2\pi J_{HS}(I_{1x}S_x + I_{1y}S_y) - \Delta\omega_{HS}S_z\} \quad [14]$$

For simplicity, we will begin by considering a field regime where the resonance frequency difference between the hydride and ligand protons $\Delta\omega_{HL}$ can be ignored, and the resonance frequency differences between the protons and heteronucleus are large enough to truncate the nonsecular term in those couplings $2\pi J_{HS}(\hat{I}_{1x}\hat{S}_x + \hat{I}_{1y}\hat{S}_y)$ and $2\pi J_{LS}(\hat{S}_x\hat{L}_x + \hat{S}_y\hat{L}_y)$. This would imply fields on the order of 10 µT-1 mT; in other field regimes there will be additional competing terms. This leaves us with

$$\hat{\mathcal{H}}' = 2\pi J_{HH}\vec{I}_1 \cdot \vec{I}_2 + 2\pi J_{HL}I_{1z}L_z - \Delta\omega_{HS}S_z + 2\pi J_{HL}(I_{1x}S_x + I_{1y}L_y) + 2\pi J_{HS}I_{1z}S_z + 2\pi J_{LS}L_zS_z \quad [15]$$

We now use equation [15] with the initial density matrix [1] and do a Taylor expansion as before for the evolution. Eq.[15-22] illustrate eight of the pathways which lead, in the fourth order, to two-spin order $L_zS_z$. It is important to note that this order lies *in the ligand itself,* and thus survives dissociation of the ligand from the transition metal complex.

$$I_{1z}I_{2z} \xrightarrow{J_{HL}(I_{1x}L_x)} iI_{1y}I_{2z}L_x \xrightarrow{J_{HH}(I_{1z}I_{2z})} -I_{1x}L_x \xrightarrow{J_{LS}(L_zS_z)} iI_{1x}L_yS_z \xrightarrow{J_{HL}(I_{1x}L_x)} L_zS_z \quad [16]$$

$$I_{1z}I_{2z} \xrightarrow{J_{HL}(I_{1x}L_x)} iI_{1y}I_{2z}L_x \xrightarrow{J_{HH}(I_{1z}I_{2z})} -I_{1x}L_x \xrightarrow{J_{HS}(I_zS_z)} iI_{1y}L_xS_z \xrightarrow{J_{HL}(I_{1y}L_y)} -L_zS_z \quad [17]$$

$$I_{1z}I_{2z} \xrightarrow{J_{HL}(I_{1x}L_x)} iI_{1y}I_{2z}L_x \xrightarrow{J_{LS}(L_zS_z)} I_{1y}I_{2z}L_yS_z \xrightarrow{J_{HH}(I_{1z}I_{2z})} iI_{1x}L_yS_z \xrightarrow{J_{HL}(I_{1x}L_x)} L_zS_z \quad [18]$$

$$I_{1z}I_{2z} \xrightarrow{J_{HL}(I_{1x}L_x)} iI_{1y}I_{2z}L_x \xrightarrow{J_{HS}(I_{1z}S_z)} -I_{1x}I_{2z}L_xS_z \xrightarrow{J_{HH}(I_{1z}I_{2z})} iI_{1y}L_xS_z \xrightarrow{J_{HL}(I_{1y}L_y)} -L_zS_z \quad [19]$$

$$I_{1z}I_{2z} \xrightarrow{J_{HL}(I_{1y}L_y)} -iI_{1x}I_{2z}L_y \xrightarrow{J_{HH}(I_{1z}I_{2z})} -I_{1y}L_y \xrightarrow{J_{LS}(L_zS_z)} -iI_{1y}L_xS_z \xrightarrow{J_{HL}(I_{1y}L_y)} L_zS_z \quad [20]$$

$$I_{1z}I_{2z} \xrightarrow{J_{HL}(I_{1y}L_y)} -iI_{1x}I_{2z}L_y \xrightarrow{J_{HH}(I_{1z}I_{2z})} -I_{1y}L_y \xrightarrow{J_{HS}(I_zS_z)} -iI_{1x}L_yS_z \xrightarrow{J_{HL}(I_{1x}L_x)} -L_zS_z \quad [21]$$

$$I_{1z}I_{2z} \xrightarrow{J_{HL}(I_{1y}L_y)} -iI_{1x}I_{2z}L_y \xrightarrow{J_{LS}(L_zS_z)} I_{1x}I_{2z}L_xS_z \xrightarrow{J_{HH}(I_{1z}I_{2z})} -iI_{1y}L_xS_z \xrightarrow{J_{HL}(I_{1y}L_y)} L_zS_z \quad [22]$$

$$I_{1z}I_{2z} \xrightarrow{J_{HL}(I_{1y}L_y)} -iI_{1x}I_{2z}L_y \xrightarrow{J_{HS}(I_{1z}S_z)} -I_{1y}I_{2z}L_yS_z \xrightarrow{J_{HH}(I_{1z}I_{2z})} -iI_{1x}L_yS_z \xrightarrow{J_{HL}(I_{1x}L_x)} -L_zS_z \quad [23]$$

As shown explicitly in the Supplementary Information, there are a total of 20 pathways which proceed from $\vec{I}_1 \cdot \vec{I}_2$ (the eight above which proceed from $I_{1z}I_{2z}$, plus six each from $I_{1x}I_{2x}$ and $I_{1y}I_{2y}$). In all of the pathways, the first step is creation of three-spin order with both hydride protons and one of the ligand nuclei. The net effect of the next two steps is to transform this into three-spin order with only the coupled hydride proton and two of the ligand nuclei, with either a two-spin or a four-spin state as the intermediate. The last step strips the hydride term. All of the many terms that can be written out are proportional to $J_{HH}J_{HL}^2$, but there are pathways proportional to either $J_{HS}$ or $J_{LS}$.

All of these pathways would have been dismissed in past work since they do not result in observable $L_z$ or $S_z$. However, the critical point we make here is that *two-spin $L_zS_z$ order in the ligand itself* can efficiently create magnetization. After a standard 90 pulse, it becomes antiphase magnetization ($L_xS_z$ or $L_zS_x$) which converts completely into in-phase magnetization $L_y$ or $S_y$ (respectively) after a time $t = 1/2J_{LS}$.

$$L_zS_z \xrightarrow{90_S, phase\, x} L_zS_y \xrightarrow{free\ evolution} L_zS_y \cos(\pi J_{LS}t) - \frac{1}{2}S_x \sin(\pi J_{LS}t)$$

$$\xrightarrow{90_L, phase\, x} L_yS_z \xrightarrow{free\ evolution} L_yS_z \cos(\pi J_{LS}t) - \frac{1}{2}L_x \sin(\pi J_{LS}t) \quad [24]$$

So, it does not matter at all that it takes a fifth coupling to convert to observable magnetization, as long as $J_{LS}$ is not vanishingly small.

A qualitative way to understand nonresonant SABRE is that the newly-added couplings $2\pi J_{HS}I_{1z}S_z + 2\pi J_{LS}L_zS_z$ act, when spin $S$ is up, to create a resonance frequency difference between spin $I_1$ and spin $L$, just as the term $\Delta\omega_{HL}$ did in the original SABRE derivation. Thus, the coupling plays the same role as the magnetic field did in the SABRE case. If $J_{HS}$-$J_{LS}$>0 and spin S is up, the hydride spin is at the higher frequency; if spin $S$ is up, the hydride spin is at the lower frequency and the produced $L$ magnetization is inverted. This means that the order is two-spin rather than one-spin.

This derivation reproduces the experimental features in Figures 3-4. It is *independent of the applied field magnitude and direction* (subject to the constraint above, that we are only considering the secular part of the $J_{LS}$ coupling). Inverting the field, or pointing it in an arbitrary direction, makes no difference. This mechanism is equally effective at producing $L$ or $S$ magnetization. In the regime where neither SABRE nor SABRE-SHEATH is efficient (e.g. $10 - 1000$ µT), this pathway will dominate; in other regimes it competes. Finally, since it arises from two-spin order, the nonresonant pathway does not share one important advantage SABRE-SHEATH has over SABRE: the ability to store magnetization on a longer-lived spin. In the usual limit, $^1$H relaxation is much faster than $^{13}$C or $^{15}$N relaxation, so in practice the two-spin order decays away with essentially the $^1$H relaxation time.

Based on these results, we can also do a more detailed interpretation of the experimental data. For our purposes, it is a good approximation to model $^{15}$N pyridine as an AX$_2$ spin system at high field, here retaining only the two ortho protons and their -10 Hz coupling to the nitrogen. Adding in labels $L_1$ and $L_2$ for the two ortho protons, we expect the resonant pathway to create $S_z$ and the nonresonant pathway to create $L_{1z}S_z + L_{2z}S_z$. Giving a 90y pulse to the nitrogen, then calculating evolution under the operators $J_{LS}(L_{1z}S_z + L_{2z}S_z)$ gives the time dependence of these two pathways as:

$$S_z \xrightarrow{90_{S,y} + free\ evolution} S_x \cos^2(\pi J_{LS}t) - 2S_y(L_{1z} + L_{2z})\cos(\pi J_{LS}t)\sin(\pi J_{LS}t) - 4S_x(L_{1z}L_{2z})\sin^2(\pi J_{LS}t) \quad [25]$$

$$S_z(L_{1z} + L_{2z}) \xrightarrow{90_{S,y} + free\ evolution} S_x(L_{1z} + L_{2z})\cos(2\pi J_{LS}t) + 2S_y L_{1z}L_{2z}\sin(2\pi J_{LS}t) + \frac{1}{2}S_x \sin(2\pi J_{LS}t) \quad [26]$$

Thus, the maximum observable signal for the nonresonant case is around 25ms ($t = 1/4J_{LS}$) as seen experimentally. At that time, the resonant SABRE-SHEATH signal is half of its initial value. This lets us combine the $t = 0$ and $t = 1/4J_{NH}$ signals to measure the relative size of one-spin and two-spin z order (Fig. 6).

The 20 pathways (in the SI) from $\vec{I}_1 \cdot \vec{I}_2$ to $L_zS_z$ include eight terms of the form $J_{HL}^2 J_{HH} J_{LS}$ (all of which lead to positive $L_zS_z$) and twelve terms of the form $J_{HL}^2 J_{HH} J_{HS}$ (six leading to positive $L_zS_z$, six to negative $L_zS_z$) so the expected functional dependence is $J_{HL}^2 J_{HH} J_{LS}$. However, our long-time simulations, which is what are relevant for generating experimental signal, show the functional dependence is $J_{HL}^2 J_{HH}(J_{LS} - J_{HS})$. Figure 7 shows an example of this. Simulations show a broad resonance creating usable magnetization with two gaps: one around $J_{LS} = 0$ (where the antiphase order cannot convert back into magnetization) and one where $J_{HS} = J_{LS}$.

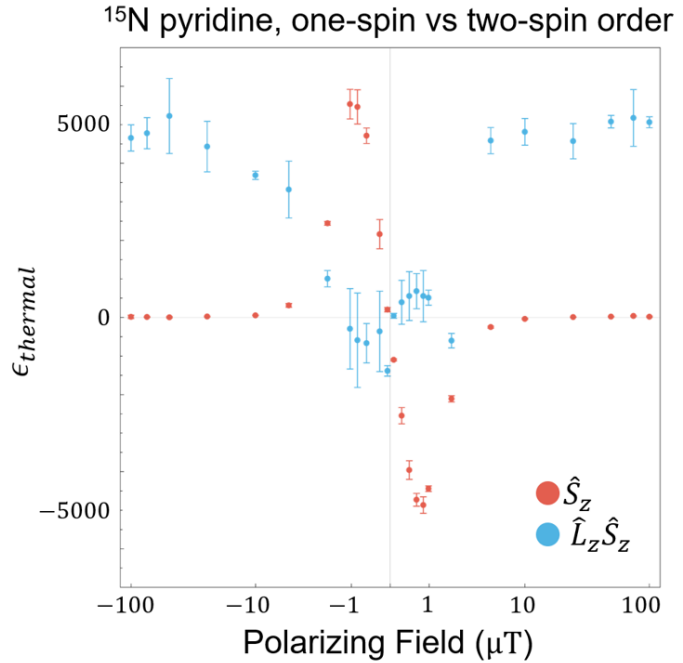

Figure 6. Inferred one-spin and two-spin order from the experimental data in Figures 2 and 3. The one-spin order can be read as the first point in the FID. The two spin order is found by subtracting half of the $t = 0$ signal from the $t = 1/4J_{NH}$ signal, then multiplying by four to account for the two L spins and the factor of ½ in equation [11]

There is a simple explanation for this discrepancy. The eight pathways from $I_{1z}I_{2z}$ to $L_zS_z$ in eq. [15-22] include four terms of the form $J_{HL}^2 J_{HH} J_{LS}$ (all of which lead to positive $L_zS_z$) and four terms of the form $J_{HL}^2 J_{HH} J_{HS}$ (all of which lead to negative $L_zS_z$), which is this exact functional dependence. This strongly suggests that the only term which matters in the initial density matrix for the long-time behavior is the $I_{1z}I_{2z}$ term, which is equivalent to a 50:50 mixture of singlet (αβ-βα) and triplet (αβ+βα)). This was verified numerically by replacing the initial condition [1] with

$$\hat{\rho}_0 = \left(\tfrac{1}{4}\hat{1} - I_{1z}I_{2z}\right) \otimes \hat{1} \qquad [25]$$

and replenishing magnetization with the same 50:50 mixture, which as Figure 7 shows has virtually no impact on the predicted signal. This is also true in normal SABRE-SHEATH (data shown in SI).

In effect, what the simulation is showing is that when ligand exchange is much more rapid than hydride exchange, the effect of the $J_{HL}I_{1z}L_z$ and $J_{HS}I_{1z}L_z$ couplings (which commute with $I_{1z}I_{2z}$ but not $I_{1x}I_{2x}$ or $I_{1y}I_{2y}$) is to convert the hydride $I_{1x}I_{2x}$ and $I_{1y}I_{2y}$ order into three-spin terms (e.g. $I_{1y}I_{2x}L_z$) which do not survive ligand dissociation. Thus, in the usual limit where the hydride dissociation is at least an order of magnitude slower than the ligand dissociation, only the $I_{1z}I_{2z}$ term is an effective hyperpolarization source most of the time. This is also a known issue at high fields due to para-ortho interconversion.[80, 81]

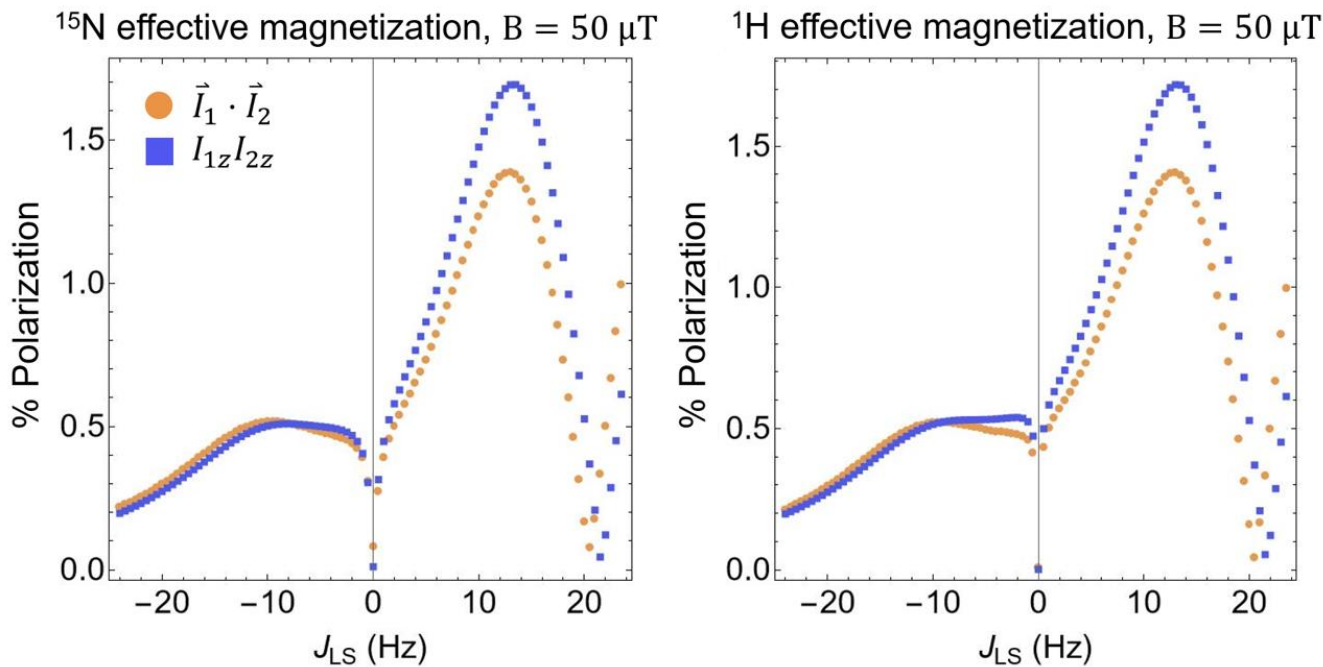

Figure 7. Comparison of the predicted efficiency of generating usable magnetization at 50 µT for a four-spin system modeled on the parameters of pyridine ($J_{HH} = -7$ Hz, $J_{HS} = 19.6$ Hz, $J_{HL} = 1.2$ Hz, $J_{LL} = -0.16$ Hz) as the parameter $J_{LS}$ is varied. The efficiency varies smoothly except around $J_{LS}=0$ (when that coupling is nearly zero, antiphase magnetization cannot convert to single-spin order) and around $J_{LS}=J_{HS}$. This agrees with the predicted $J_{HL}^2 J_{HH}(J_{LS} - J_{HS})$ dependence starting from $I_{1z}I_{2z}$. The orange curves assume the catalyst starts with singlet hydrides and is always replenished with singlet; the blue represents an extreme case where the hydrides and hydrogen in solution have only $I_{1z}I_{2z}$ order. The signals are nearly identical (with a slight shift of the zero, reflecting some action by a mechanism proportional to $J_{HL}^2 J_{HH} J_{LS}$ from the $I_{1y}I_{2y}$ and $I_{1z}I_{2x}$ order). Simulations at the SABRE-SHEATH peak (0.5 µT) also show little effects from $I_{1y}I_{2y}$ and $I_{1x}I_{2x}$ order. This insensitivity comes because, in the usual limit where hydrides exchange far slower than ligands, $I_{1y}I_{2y}$ and $I_{1x}I_{2x}$ order is quenched almost all the time (as discussed in text).

Discussion

Inclusion of the additional pathways that create two-spin magnetization has several practical consequences which have not yet been fully explored but are likely meaningful for SABRE applications.

—The pathways are independent of magnetic field over a broad range, but additional terms come in at frequencies lower than the SABRE-SHEATH or higher than the SABRE condition. This will lead to additional optimal frequencies, particularly in the SABRE case. For example, the C-H coupling in a methyl group $^{13}CH_3$ is typically about 130 Hz. This implies that if the carbon is spin-down, the methyl protons have an intrinsic 130 Hz difference between spin up and spin down, which is the same as the frequency difference those protons would have from the hydride protons at 120 mT (assuming 25 ppm chemical shift difference). In the normal SABRE condition, the 25 ppm frequency difference between the hydride and methyl protons gives about a 7 Hz frequency difference at 6.5 mT. But this suggests a new resonance would exist offset by this amount from 120 mT, where this field cancels out the $J_{CH}$ coupling. In this case, spin-down $^{13}C$ nuclei allow polarization of $^1H$ (because the ligand-hydride resonance frequency difference is matched to the traditional SABRE condition) but spin-up $^{13}C$ nuclei produce a large frequency difference so they are inactive. Mathematically, this implies normal $^1H$ and antiphase $^1H$-$^{13}C$ magnetization are produced in equal quantities. This resonance will move with the chemical shift difference (so it provides some selectivity), and is near the range of commercially available low-field MRI scanners[74]. This effect is also likely responsible for anomalous observations of SABRE resonances near 100 mT in $^{13}$C labeled pyridines[66].

—While we have only explicitly considered the case of $^1$H plus a heteronucleus, there are likely extensions to two heteronuclei. One example is the $^{15}$N-$^{77}$Se case which has recently been explored experimentally[82] and to direct pumping of homonuclear singlets such as in diazirine[83, 84]. In the latter case, direct fourth-order pathways from hydride singlet order to ligand singlet order (e.g. $\vec{I}_1 \bullet \vec{I}_2 \xrightarrow{J_{HL}} I_{1z}I_{2y}L_x(cyc) \xrightarrow{J_{HH}} \vec{I}_1 \bullet \vec{L} \xrightarrow{J_{LS}} I_{1x}L_yS_z(cyc) \xrightarrow{J_{HL}} \vec{S} \bullet \vec{L}$) is also field independent.

—Most of the terms in the Hamiltonian in equation [14] are readily controlled, inverted or cancelled by radiofrequency pulses, in the standard ways known in NMR for over half a century. This provides a different strategy for doing high-field SABRE than has been commonly explored. For example, again thinking in terms of field strengths around 100 mT, it would be straightforward to zero out the terms from equation [14] that we omitted in equation [15] using multiple echo sequences on both nuclei, and thus to extend the range of operation of the nonresonant SABRE effect.

—There are likely significant advantages to this strategy for high field SABRE experiments, because most existing methods are highly sensitive to the population difference between hydride singlet (αβ-βα) and the triplet state (αβ+βα). For example, the first such experiment (LIGHT-SABRE)[85] showed explicitly that polarization is created by weak irradiation at the nitrogen frequency, interchanging states with those two different hydride characters. As noted earlier, the fragility of the $I_{1x}I_{2x}$ and $I_{1y}I_{2y}$ order implies a strong tendency to equalize those populations ($I_{1z}I_{2z}$ has equal population of these two states), which cancels the LIGHT-SABRE effect. Thus, nonresonant SABRE works fine with $I_{1z}I_{2z}$ order, whereas most existing high field SABRE approaches do not. This has been addressed in some recent papers[81, 86, 87] with relatively complex pulse sequences, but using the nonresonant mechanism will likely improve pulse sequence strategies.

—Finally, extensive optimization of SABRE-SHEATH approaches by computer has led to robust, three-dimensional fields that are accessible in the laboratory[88]. To our knowledge, none of these calculations has ever tried to optimize anything other than magnetization. Feeding two-spin as well as one-spin order into optimization approaches is likely to improve results, as the pathways are largely independent of one another.

Conclusions

We have demonstrated a hitherto unappreciated para-hydrogen based hyperpolarization mechanism which involves direct creation of two-spin *zz* order in the target ligand. While here we have presented data mostly for pyridine, the mechanism should be accessible for essentially every molecule that has been hyperpolarized by SABRE and has a heteroatom near the hydrogen atoms, and we have found experimental signals in other common molecules such as acetonitrile (see SI). One reason this may have gone undetected until now is relatively simple. In normal hyperpolarization experiments, magnetization is detected by the signal directly after a 90 pulse, so inhomogeneous broadening in the spectrometer is not important. Thus, SABRE testing is often done without removing the bubbling apparatus (which causes susceptibility effects). This is a critical problem for detecting the antiphase magnetization in nonlinear SABRE, particularly if (unlike the pyridine case) the coupling $J_{LS}$ is small, as the signal disappears before it can refocus. This is of course easily solved in the laboratory — it was just that it was not realized it might be important.

The nonresonant mechanism has multiple desirable features. These include near-independence of external field strength over multiple orders of magnitude (in fact, it can be done quite well in an uncontrolled lab setting in the Earth's field) and low sensitivity to the coupling-induced interconversion between hydride singlet and triplet order. This mechanism likely provides multiple new strategies for SABRE optimization at a wide variety of fields. This amplifies the advantages of SABRE (cheap and simple source of spin order, simple and low-cost apparatus) as the current best candidate for a truly general and versatile hyperpolarization strategy.

This work is supported by the National Science Foundation under grant CHE-2505898.

1. Hirsch ML, Kalechofsky N, Belzer A, Rosay M, Kempf JG. Brute-Force Hyperpolarization for NMR and MRI. Journal of the American Chemical Society. 2015;137(26):8428-34. doi: 10.1021/jacs.5b01252.
2. Waugh JS, Hammel PG, Kuhns PL, Gonen O. Spin-Lattice Relaxation below 1K. Bulletin of Magnetic Resonance. 1989;11:97-102.
3. Zangara PR, Dhomkar S, Ajoy A, Liu K, Nazaryan R, Pagliero D, Suter D, Reimer JA, Pines A, Meriles CA. Dynamics of frequency-swept nuclear spin optical pumping in powdered diamond at low magnetic fields. Proceedings of the National Academy of Sciences. 2019;116(7):2512-20. doi: 10.1073/pnas.1811994116.
4. Salerno M, Altes TA, Mugler JP, 3rd, Nakatsu M, Hatabu H, de Lange EE. Hyperpolarized noble gas MR imaging of the lung: potential clinical applications. European journal of radiology. 2001;40(1):33-44. PubMed PMID: 11673006.
5. MacFall JR, Charles HC, Black RD, Middleton H, Swartz JC, Saam B, Driehuys B, Erickson C, Happer W, Cates GD, Johnson GA, Ravin CE. Human lung air spaces: potential for MR imaging with hyperpolarized He-3. Radiology. 1996;200(2):553-8. PubMed PMID: 8685356.
6. Branca RT, Cleveland ZI, Fubara B, Kumar CS, Maronpot RR, Leuschner C, Warren WS, Driehuys B. Molecular MRI for sensitive and specific detection of lung metastases. Proc Natl Acad Sci U S A. 2010;107(8):3693-7. Epub 2010/02/10. doi: 10.1073/pnas.1000386107. PubMed PMID: 20142483; PMCID: 2840507.
7. Pavlovskaya GE, Cleveland ZI, Stupic KF, Basaraba RJ, Meersmann T. Hyperpolarized krypton-83 as a contrast agent for magnetic resonance imaging. Proc Natl Acad Sci U S A. 2005;102(51):18275-9. Epub 2005/12/12. doi: 10.1073/pnas.0509419102. PubMed PMID: 16344474.
8. Hersman FW, Ruset IC, Ketel S, Muradian I, Covrig SD, Distelbrink J, Porter W, Watt D, Ketel J, Brackett J, Hope A, Patz S. Large production system for hyperpolarized 129Xe for human lung imaging studies. Academic radiology. 2008;15(6):683-92. doi: 10.1016/j.acra.2007.09.020. PubMed PMID: 18486005.
9. Ebner L, Kammerman J, Driehuys B, Schiebler ML, Cadman RV, Fain SB. The role of hyperpolarized (129)xenon in MR imaging of pulmonary function. European journal of radiology. 2017;86:343-52. Epub 2016/09/16. doi: 10.1016/j.ejrad.2016.09.015. PubMed PMID: 27707585.
10. Henstra A, Lin TS, Schmidt J, Wenckebach WT. High dynamic nuclear polarization at room temperature. Chemical Physics Letters. 1990;165(1):6-10. doi: 10.1016/0009-2614(90)87002-9.
11. Kouno H, Kawashima Y, Tateishi K, Uesaka T, Kimizuka N, Yanai N. Nonpentacene Polarizing Agents with Improved Air Stability for Triplet Dynamic Nuclear Polarization at Room Temperature. The Journal of Physical Chemistry Letters. 2019;10(9):2208-13. doi: 10.1021/acs.jpclett.9b00480.
12. Okuno Y, Cavagnero S. Effect of heavy atoms on photochemically induced dynamic nuclear polarization in liquids. Journal of Magnetic Resonance. 2018;286:172-87. doi: 10.1016/j.jmr.2017.12.001.
13. Hore PJ, Kaptein R. Photochemically Induced Dynamic Nuclear Polarization (Photo-CIDNP) of Biological Molecules Using Continuous Wave and Time-Resolved Methods.  NMR Spectroscopy: New Methods and Applications: AMERICAN CHEMICAL SOCIETY; 1982. p. 285-318.
14. Okuno Y, Cavagnero S. Fluorescein: A Photo-CIDNP Sensitizer Enabling Hypersensitive NMR Data Collection in Liquids at Low Micromolar Concentration. The Journal of Physical Chemistry B. 2016;120(4):715-23. doi: 10.1021/acs.jpcb.5b12339.
15. Morozova OB, Ivanov KL. Time-Resolved Chemically Induced Dynamic Nuclear Polarization of Biologically Important Molecules2019;20(2):197-215. doi: 10.1002/cphc.201800566.

16.	de Boer W, Niinikoski TO. Dynamic proton polarization in propanediol below 0.5 K. Nuclear Instruments and Methods. 1974;114(3):495-8. doi: 10.1016/0029-554X(74)90172-4.
17.	Boer W, Borghini M, Morimoto K, Niinikoski TO, Udo F. Dynamic polarization of protons, deuterons, and carbon-13 nuclei: Thermal contact between nuclear spins and an electron spin-spin interaction reservoir. J Low Temp Phys. 1974;15(3-4):249-67. doi: 10.1007/BF00661185.
18.	Bajaj VS, Farrar CT, Hornstein MK, Mastovsky I, Vieregg J, Bryant J, Eléna B, Kreischer KE, Temkin RJ, Griffin RG. Dynamic nuclear polarization at 9T using a novel 250 GHz gyrotron microwave source. Journal of Magnetic Resonance. 2003;160(2):85-90. doi: 10.1016/S1090-7807(02)00192-1.
19.	Hall DA, Maus DC, Gerfen GJ, Inati SJ, Becerra LR, Dahlquist FW, Griffin RG. Polarization-Enhanced NMR Spectroscopy of Biomolecules in Frozen Solution. Science. 1997;276(5314):930-2. doi: 10.1126/science.276.5314.930.
20.	Abragam A, Goldman M. Principles of dynamic nuclear polarisation. Reports on Progress in Physics. 1978;41(3):395.
21.	Ardenkjær-Larsen JH, Fridlund B, Gram A, Hansson G, Hansson L, Lerche MH, Servin R, Thaning M, Golman K. Increase in signal-to-noise ratio of > 10,000 times in liquid-state NMR. Proceedings of the National Academy of Sciences. 2003;100(18):10158-63. doi: 10.1073/pnas.1733835100.
22.	Abragam A, Proctor WG. Experiments on Spin Temperature. Physical Review. 1957;106(1):160-1. doi: 10.1103/PhysRev.106.160.
23.	Goertz ST. The dynamic nuclear polarization process. Nuclear Instruments and Methods in Physics Research Section A: Accelerators, Spectrometers, Detectors and Associated Equipment. 2004;526(1):28-42. doi: 10.1016/j.nima.2004.03.147.
24.	Eichhorn TR, Takado Y, Salameh N, Capozzi A, Cheng T, Hyacinthe J-N, Mishkovsky M, Roussel C, Comment A. Hyperpolarization without persistent radicals for in vivo real-time metabolic imaging. Proc Natl Acad Sci U S A. 2013;110(45):18064-9. Epub 2013/10/21. doi: 10.1073/pnas.1314928110. PubMed PMID: 24145405.
25.	Hu K-N, Debelouchina GT, Smith AA, Griffin RG. Quantum mechanical theory of dynamic nuclear polarization in solid dielectrics2011;134(12):125105. doi: 10.1063/1.3564920.
26.	Kurhanewicz J, Vigneron DB, Brindle K, Chekmenev EY, Comment A, Cunningham CH, Deberardinis RJ, Green GG, Leach MO, Rajan SS, Rizi RR, Ross BD, Warren WS, Malloy CR. Analysis of cancer metabolism by imaging hyperpolarized nuclei: prospects for translation to clinical research. Neoplasia. 2011;13(2):81-97. Epub 2011/03/16. PubMed PMID: 21403835; PMCID: 3033588.
27.	Brindle KM, Bohndiek SE, Gallagher FA, Kettunen MI. Tumor Imaging Using Hyperpolarized (13)C Magnetic Resonance. Magnetic Resonance in Medicine. 2011;66(2):505-19. doi: 10.1002/mrm.22999.
28.	Viale A, Aime S. Current concepts on hyperpolarized molecules in MRI. Current Opinion in Chemical Biology. 2010;14(1):90-6. doi: 10.1016/j.cbpa.2009.10.021.
29.	Keshari KR, Wilson DM. Chemistry and biochemistry of 13C hyperpolarized magnetic resonance using dynamic nuclear polarization. Chemical Society Reviews. 2014;43(5):1627-59. doi: 10.1039/C3CS60124B.
30.	Wigh Lipso K, Hansen ESS, Tougaard RS, Laustsen C, Ardenkjaer-Larsen JH. Renal MR angiography and perfusion in the pig using hyperpolarized water. Magn Reson Med. 2017;78(3):1131-5. Epub 2016/10/01. doi: 10.1002/mrm.26478. PubMed PMID: 27690362.
31.	Sadet A, Stavarache C, Bacalum M, Radu M, Bodenhausen G, Kurzbach D, Vasos PR. Hyperpolarized Water Enhances Two-Dimensional Proton NMR Correlations: A New Approach for

Molecular Interactions. J Am Chem Soc. 2019;141(32):12448-52. Epub 2019/08/02. doi: 10.1021/jacs.9b03651. PubMed PMID: 31368708.
32. Olsen G, Markhasin E, Szekely O, Bretschneider C, Frydman L. Optimizing water hyperpolarization and dissolution for sensitivity-enhanced 2D biomolecular NMR. Journal of magnetic resonance (San Diego, Calif : 1997). 2016;264:49-58. Epub 2016/02/28. doi: 10.1016/j.jmr.2016.01.005. PubMed PMID: 26920830.
33. Lipso KW, Bowen S, Rybalko O, Ardenkjaer-Larsen JH. Large dose hyperpolarized water with dissolution-DNP at high magnetic field. Journal of magnetic resonance (San Diego, Calif : 1997). 2017;274:65-72. Epub 2016/11/28. doi: 10.1016/j.jmr.2016.11.008. PubMed PMID: 27889650.
34. Kim J, Mandal R, Hilty C. Observation of Fast Two-Dimensional NMR Spectra during Protein Folding Using Polarization Transfer from Hyperpolarized Water. J Phys Chem Lett. 2019:5463-7. Epub 2019/08/24. doi: 10.1021/acs.jpclett.9b02197. PubMed PMID: 31442055.
35. Nelson SJ, Kurhanewicz J, Vigneron DB, Larson PEZ, Harzstark AL, Ferrone M, van Criekinge M, Chang JW, Bok R, Park I, Reed G, Carvajal L, Small EJ, Munster P, Weinberg VK, Ardenkjaer-Larsen JH, Chen AP, Hurd RE, Odegardstuen L-I, Robb FJ, Tropp J, Murray JA. Metabolic Imaging of Patients with Prostate Cancer Using Hyperpolarized [1-13C]Pyruvate. Science Translational Medicine. 2013;5(198):198ra08. doi: 10.1126/scitranslmed.3006070.
36. Gallagher FA, Kettunen MI, Day SE, Hu DE, Ardenkjaer-Larsen JH, Zandt R, Jensen PR, Karlsson M, Golman K, Lerche MH, Brindle KM. Magnetic resonance imaging of pH in vivo using hyperpolarized 13C-labelled bicarbonate. Nature. 2008;453(7197):940-3. Epub 2008/05/30. doi: 10.1038/nature07017. PubMed PMID: 18509335.
37. Gallagher FA, Kettunen MI, Brindle KM. Imaging pH with hyperpolarized 13C. NMR in biomedicine. 2011;24(8):1006-15. Epub 2011/08/04. doi: 10.1002/nbm.1742. PubMed PMID: 21812047.
38. Brindle K. Watching tumours gasp and die with MRI: the promise of hyperpolarised 13C MR spectroscopic imaging. The British journal of radiology. 2012;85(1014):697-708. Epub 2012/04/13. doi: 10.1259/bjr/81120511. PubMed PMID: 22496072; PMCID: PMC3474112.
39. Chapman B, Joalland B, Meersman C, Ettedgui J, Swenson RE, Krishna MC, Nikolaou P, Kovtunov KV, Salnikov OG, Koptyug IV, Gemeinhardt ME, Goodson BM, Shchepin RV, Chekmenev EY. Low-Cost High-Pressure Clinical-Scale 50% Parahydrogen Generator Using Liquid Nitrogen at 77 K. Analytical chemistry. 2021;93(24):8476-83. doi: 10.1021/acs.analchem.1c00716.
40. Mhaske Y, Sutter E, Daley J, Mahoney C, Whiting N. 65% Parahydrogen from a liquid nitrogen cooled generator. Journal of Magnetic Resonance. 2022;341:107249. doi: 10.1016/j.jmr.2022.107249.
41. Sundararajan K, Sankaran K, Ramanathan N, Gopi R. Production and characterization of para-hydrogen gas for matrix isolation infrared spectroscopy. Journal of Molecular Structure. 2016;1117:181-91. doi: 10.1016/j.molstruc.2016.03.068.
42. Silvera IF. The solid molecular hydrogens in the condensed phase: Fundamentals and static properties. Reviews of Modern Physics. 1980;52(2):393-452. doi: 10.1103/RevModPhys.52.393.
43. Das T, Kweon S-C, Nah IW, Karng SW, Choi J-G, Oh I-H. Spin conversion of hydrogen using supported iron catalysts at cryogenic temperature. Cryogenics. 2015;69:36-43. doi: 10.1016/j.cryogenics.2015.03.003.
44. Paris S, Ilisca E. Electron−Nucleus Resonances and Magnetic Field Accelerations in the Ortho−Para H2 Conversion. The Journal of Physical Chemistry A. 1999;103(25):4964-8. doi: 10.1021/jp990040t.
45. Wagner S. Conversion rate of para-hydrogen to ortho-hydrogen by oxygen: implications for PHIP gas storage and utilization. Magn Reson Mater Phy. 2014;27(3):195-9. doi: 10.1007/s10334-013-0399-y.

46. Golman K, Axelsson O, Jóhannesson H, Månsson S, Olofsson C, Petersson JS. Parahydrogen-induced polarization in imaging: Subsecond 13C angiography. Magnetic Resonance in Medicine. 2001;46(1):1-5. doi: 10.1002/mrm.1152.
47. Bowers CR, Weitekamp DP. Parahydrogen and synthesis allow dramatically enhanced nuclear alignment. Journal of the American Chemical Society. 1987;109(18):5541-2. doi: 10.1021/ja00252a049.
48. Natterer J, Bargon J. Parahydrogen induced polarization. Prog Nucl Mag Res Sp. 1997;31:293-315. doi: 10.1016/S0079-6565(97)00007-1.
49. Duckett SB, Sleigh CJ. Applications of the parahydrogen phenomenon: A chemical perspective. Prog Nucl Mag Res Sp. 1999;34(1):71-92. doi: 10.1016/S0079-6565(98)00027-2.
50. Bowers CR, Weitekamp DP. Transformation of Symmetrization Order to Nuclear-Spin Magnetization by Chemical-Reaction and Nuclear-Magnetic-Resonance. Phys Rev Lett. 1986;57(21):2645-8. doi: 10.1103/PhysRevLett.57.2645.
51. Eisenschmid TC, Kirss RU, Deutsch PP, Hommeltoft SI, Eisenberg R, Bargon J, Lawler RG, Balch AL. Para hydrogen induced polarization in hydrogenation reactions. Journal of the American Chemical Society. 1987;109(26):8089-91. doi: 10.1021/ja00260a026.
52. Adams RW, Duckett SB, Green RA, Williamson DC, Green GGR. A theoretical basis for spontaneous polarization transfer in non-hydrogenative parahydrogen-induced polarization. J Chem Phys. 2009;131(19).
53. Atkinson KD, Cowley MJ, Elliott PI, Duckett SB, Green GG, Lopez-Serrano J, Whitwood AC. Spontaneous transfer of parahydrogen derived spin order to pyridine at low magnetic field. J Am Chem Soc. 2009;131(37):13362-8. Epub 2009/09/02. doi: 10.1021/ja903601p. PubMed PMID: 19719167.
54. Duckett SB, Mewis RE. Improving NMR and MRI Sensitivity with Parahydrogen. Topics in current chemistry. 2012. Epub 2012/11/10. doi: 10.1007/128_2012_388. PubMed PMID: 23138689.
55. Duckett SB, Mewis RE. Application of parahydrogen induced polarization techniques in NMR spectroscopy and imaging. Accounts of chemical research. 2012;45(8):1247-57. Epub 2012/03/29. doi: 10.1021/ar2003094. PubMed PMID: 22452702.
56. Adams RW, Aguilar JA, Atkinson KD, Cowley MJ, Elliott PI, Duckett SB, Green GG, Khazal IG, López-Serrano J, Williamson DC. Reversible interactions with para-hydrogen enhance NMR sensitivity by polarization transfer. Science. 2009;323(5922):1708-11.
57. Theis T, Truong ML, Coffey AM, Shchepin RV, Waddell KW, Shi F, Goodson BM, Warren WS, Chekmenev EY. Microtesla SABRE enables 10% nitrogen-15 nuclear Spin polarization. Journal of the American Chemical Society. 2015;137(4):1404-7. doi: 10.1021/ja512242d.
58. Svyatova A, Skovpin IV, Chukanov NV, Kovtunov KV, Chekmenev EY, Pravdivtsev AN, Hovener JB, Koptyug IV. (15) N MRI of SLIC-SABRE Hyperpolarized (15) N-Labelled Pyridine and Nicotinamide. Chemistry (Weinheim an der Bergstrasse, Germany). 2019;25(36):8465-70. Epub 2019/04/06. doi: 10.1002/chem.201900430. PubMed PMID: 30950529.
59. Stepanek P, Sanchez-Perez C, Telkki VV, Zhivonitko VV, Kantola AM. High-throughput continuous-flow system for SABRE hyperpolarization. Journal of magnetic resonance (San Diego, Calif : 1997). 2019;300:8-17. Epub 2019/01/27. doi: 10.1016/j.jmr.2019.01.003. PubMed PMID: 30684826.
60. Stepanek P, Kantola AM. Low-Concentration Measurements of Nuclear Spin-Induced Optical Rotation Using SABRE Hyperpolarization. J Phys Chem Lett. 2019;10(18):5458-62. Epub 2019/08/28. doi: 10.1021/acs.jpclett.9b02194. PubMed PMID: 31454246.
61. Richardson PM, John RO, Parrott AJ, Rayner PJ, Iali W, Nordon A, Halse ME, Duckett SB. Quantification of hyperpolarisation efficiency in SABRE and SABRE-Relay enhanced NMR spectroscopy. Physical chemistry chemical physics : PCCP. 2018;20(41):26362-71. Epub 2018/10/12. doi: 10.1039/c8cp05473h. PubMed PMID: 30303501; PMCID: PMC6202922.

62. Kovtunov KV, Pokochueva EV, Salnikov OG, Cousin SF, Kurzbach D, Vuichoud B, Jannin S, Chekmenev EY, Goodson BM, Barskiy DA, Koptyug IV. Hyperpolarized NMR Spectroscopy: d-DNP, PHIP, and SABRE Techniques. Chemistry, an Asian journal. 2018. Epub 2018/05/24. doi: 10.1002/asia.201800551. PubMed PMID: 29790649; PMCID: PMC6251772.
63. Buckenmaier K, Scheffler K, Plaumann M, Fehling P, Bernarding J, Rudolph M, Back C, Kolle D, Kleiner R, Hovener JB, Pravdivtsev A. Hyperpolarized high order multiple quantum coherences at ultra-low fields. Chemphyschem : a European journal of chemical physics and physical chemistry. 2019. Epub 2019/09/20. doi: 10.1002/cphc.201900757. PubMed PMID: 31536665.
64. Pravdivtsev AN, Skovpin IV, Svyatova AI, Chukanov NV, Kovtunova LM, Bukhtiyarov VI, Chekmenev EY, Kovtunov KV, Koptyug IV, Hövener J-B. Chemical Exchange Reaction Effect on Polarization Transfer Efficiency in SLIC-SABRE. The Journal of Physical Chemistry A. 2018;122(46):9107-14. doi: 10.1021/acs.jpca.8b07163.
65. Pravdivtsev AN, Yurkovskaya AV, Vieth H-M, Ivanov KL. RF-SABRE: A Way to Continuous Spin Hyperpolarization at High Magnetic Fields. The Journal of Physical Chemistry B. 2015;119(43):13619-29. doi: 10.1021/acs.jpcb.5b03032.
66. Kiryutin AS, Yurkovskaya AV, Zimmermann H, Vieth HM, Ivanov KL. Complete magnetic field dependence of SABRE-derived polarization. Magnetic resonance in chemistry : MRC. 2018;56(7):651-62. Epub 2017/12/13. doi: 10.1002/mrc.4694. PubMed PMID: 29230864.
67. Tickner BJ, Dennington M, Collins BG, Gater CA, Tanner TFN, Whitwood AC, Rayner PJ, Watts DP, Duckett SB. Metal-Mediated Catalytic Polarization Transfer from para Hydrogen to 3,5-Dihalogenated Pyridines. ACS Catal. 2024;14(2):994-1004. Epub 20240105. doi: 10.1021/acscatal.3c05378. PubMed PMID: 38269038; PMCID: PMC10804365.
68. Them K, Kuhn J, Pravdivtsev AN, Hövener JB. Nuclear spin polarization of lactic acid via exchange of parahydrogen-polarized protons. Commun Chem. 2024;7(1):172. Epub 20240808. doi: 10.1038/s42004-024-01254-8. PubMed PMID: 39112677; PMCID: PMC11306230.
69. Salnikov OG, Assaf CD, Yi AP, Duckett SB, Chekmenev EY, Hövener JB, Koptyug IV, Pravdivtsev AN. Modeling Ligand Exchange Kinetics in Iridium Complexes Catalyzing SABRE Nuclear Spin Hyperpolarization. Analytical chemistry. 2024;96(29):11790-9. Epub 20240708. doi: 10.1021/acs.analchem.4c01374. PubMed PMID: 38976810; PMCID: PMC11270526.
70. Robinson AD, Hill-Casey F, Duckett SB, Halse ME. Quantitative reaction monitoring using parahydrogen-enhanced benchtop NMR spectroscopy. Physical chemistry chemical physics : PCCP. 2024;26(19):14317-28. Epub 20240515. doi: 10.1039/d3cp06221j. PubMed PMID: 38695736.
71. Nantogma S, de Maissin H, Adelabu I, Abdurraheem A, Nelson C, Chukanov NV, Salnikov OG, Koptyug IV, Lehmkuhl S, Schmidt AB, Appelt S, Theis T, Chekmenev EY. Carbon-13 Radiofrequency Amplification by Stimulated Emission of Radiation of the Hyperpolarized Ketone and Hemiketal Forms of Allyl [1-(13C)]Pyruvate. ACS Sens. 2024;9(2):770-80. Epub 20240110. doi: 10.1021/acssensors.3c02075. PubMed PMID: 38198709; PMCID: PMC10922715.
72. Nantogma S, Chowdhury MRH, Kabir MSH, Adelabu I, Joshi SM, Samoilenko A, de Maissin H, Schmidt AB, Nikolaou P, Chekmenev YA, Salnikov OG, Chukanov NV, Koptyug IV, Goodson BM, Chekmenev EY. MATRESHCA: Microtesla Apparatus for Transfer of Resonance Enhancement of Spin Hyperpolarization via Chemical Exchange and Addition. Analytical chemistry. 2024;96(10):4171-9. Epub 20240215. doi: 10.1021/acs.analchem.3c05233. PubMed PMID: 38358916; PMCID: PMC10939749.
73. Kempf N, Körber R, Plaumann M, Pravdivtsev AN, Engelmann J, Boldt J, Scheffler K, Theis T, Buckenmaier K. (13)C MRI of hyperpolarized pyruvate at 120 µT. Sci Rep. 2024;14(1):4468. Epub 20240223. doi: 10.1038/s41598-024-54770-x. PubMed PMID: 38396023; PMCID: PMC10891046.

74. Iqbal N, Brittin DO, Daluwathumullagamage PJ, Alam MS, Senanayake IM, Gafar AT, Siraj Z, Petrilla A, Pugh M, Tonazzi B, Ragunathan S, Poorman ME, Sacolick L, Theis T, Rosen MS, Chekmenev EY, Goodson BM. Toward Next-Generation Molecular Imaging with a Clinical Low-Field (0.064 T) Point-of-Care MRI Scanner. Analytical chemistry. 2024;96(25):10348-55. Epub 20240610. doi: 10.1021/acs.analchem.4c01299. PubMed PMID: 38857182.
75. Goodson BM, Chekmenev EY. Toward next-generation molecular imaging. Proc Natl Acad Sci U S A. 2024;121(18):e2405380121. Epub 20240424. doi: 10.1073/pnas.2405380121. PubMed PMID: 38657055; PMCID: PMC11067020.
76. Ettedgui J, Yamamoto K, Blackman B, Koyasu N, Raju N, Vasalatiy O, Merkle H, Chekmenev EY, Goodson BM, Krishna MC, Swenson RE. In vivo Metabolic Sensing of Hyperpolarized [1-(13)C]Pyruvate in Mice Using a Recyclable Perfluorinated Iridium Signal Amplification by Reversible Exchange Catalyst. Angewandte Chemie (International ed in English). 2024:e202407349. Epub 20240603. doi: 10.1002/anie.202407349. PubMed PMID: 38829568.
77. Ettedgui J, Blackman B, Raju N, Kotler SA, Chekmenev EY, Goodson BM, Merkle H, Woodroofe CC, LeClair CA, Krishna MC, Swenson RE. Perfluorinated Iridium Catalyst for Signal Amplification by Reversible Exchange Provides Metal-Free Aqueous Hyperpolarized [1-(13)C]-Pyruvate. J Am Chem Soc. 2024;146(1):946-53. Epub 20231228. doi: 10.1021/jacs.3c11499. PubMed PMID: 38154120; PMCID: PMC10785822.
78. Eriksson SL, Lindale JR, Li X, Warren WS. Improving SABRE hyperpolarization with highly nonintuitive pulse sequences: Moving beyond avoided crossings to describe dynamics. Science advances. 2022;8(11):eabl3708. Epub 20220316. doi: 10.1126/sciadv.abl3708. PubMed PMID: 35294248; PMCID: PMC8926330.
79. Lindale JR, Eriksson SL, Tanner CPN, Warren WS. Infinite-order perturbative treatment for quantum evolution with exchange. Science advances. 2020;6(32):eabb6874. Epub 2020/08/22. doi: 10.1126/sciadv.abb6874. PubMed PMID: 32821841; PMCID: PMC7413723.
80. Snadin AV, Chuklina NO, Kiryutin AS, Lukzen NN, Yurkovskaya AV. Magnetic field dependence of the para-ortho conversion rate of molecular hydrogen in SABRE experiments. Journal of Magnetic Resonance. 2024;360:107630. doi: https://doi.org/10.1016/j.jmr.2024.107630.
81. Markelov DA, Kozinenko VP, Knecht S, Kiryutin AS, Yurkovskaya AV, Ivanov KL. Singlet to triplet conversion in molecular hydrogen and its role in parahydrogen induced polarization. Physical chemistry chemical physics : PCCP. 2021;23(37):20936-44. Epub 20210929. doi: 10.1039/d1cp03164c. PubMed PMID: 34542122.
82. Kiryutin AS, Markelov DA, Matsulevich ZV, Kosenko ID, Kireev NV, Godovikov IA, Yurkovskaya AV. Microtesla Signal Amplification by Reversible Exchange Enables Simultaneous over 5% Polarization of 77Se and 15N at Natural Abundance in a Selenium–Nitrogen Heterocycle. Journal of the American Chemical Society. 2025;147(26):23113-9. doi: 10.1021/jacs.5c06450.
83. Theis T, Ortiz GX, Logan AW, Claytor KE, Feng Y, Huhn WP, Blum V, Malcolmson SJ, Chekmenev EY, Wang Q. Direct and cost-efficient hyperpolarization of long-lived nuclear spin states on universal 15N2-diazirine molecular tags. Science advances. 2016;2(3):e1501438.
84. Park H, Zhang G, Bae J, Theis T, Warren WS, Wang Q. Application of 15N2-Diazirines as a Versatile Platform for Hyperpolarization of Biological Molecules by d-DNP. Bioconjugate Chemistry. 2020;31(3):537-41. doi: 10.1021/acs.bioconjchem.0c00028.
85. Theis T, Truong M, Coffey AM, Chekmenev EY, Warren WS. LIGHT-SABRE enables efficient in-magnet catalytic hyperpolarization. Journal of magnetic resonance (San Diego, Calif : 1997). 2014;248:23-6.

86. Lindale JR, Eriksson SL, Warren WS. Phase coherent excitation of SABRE permits simultaneous hyperpolarization of multiple targets at high magnetic field. Phys Chem Chem Phys. 2022;24(12):7214-23. Epub 20220323. doi: 10.1039/d1cp05962a. PubMed PMID: 35266466.
87. Markelov DA, Kozinenko VP, Kiryutin AS, Yurkovskaya AV. High-field SABRE pulse sequence design for chemically non-equivalent spin systems. The Journal of Chemical Physics. 2024;161(21). doi: 10.1063/5.0236841.
88. Lindale JR, Smith LL, Mammen MW, Eriksson SL, Everhart LM, Warren WS. Multi-axis fields boost SABRE hyperpolarization. Proc Natl Acad Sci U S A. 2024;121(14):e2400066121. Epub 20240327. doi: 10.1073/pnas.2400066121. PubMed PMID: 38536754; PMCID: PMC10998558.

**Supplementary Information**
 **A. Materials and methods for experimental demonstration of nonresonant SABRE:**

$^{15}$N-pyridine SABRE samples were prepared with 160 mM $^{15}$N-pyridine (Cambridge Isotope Laboratories NLM 305-0.5, CAS 34322-45-7) and 10 mM Ir-Imes catalyst (IMes = 1,3-bis(2,4,6-trimethylphenyl)imidazol-2-ylidene, synthesized by collaborator) in methanol-d$_4$ (Cambridge Isotope Laboratories DLM-24-50, CAS 811-98-3) solvent. For the representative spectra shown in Fig. 3, samples additionally contained 10% v/v methanol (Sigma-Aldrich 34860-4L-R, CAS 67-56-1) so that the measurement magnet could be shimmed on the sample to help reduce inhomogeneity resulting from the presence of the H$_2$ gas bubbler in the NMR tube. The sample tube was a Wilmad 5mm sapphire NMR tube (WG-507-7) custom-fitted to adapt to the H$_2$ gas line which was pressurized to 120 PSI.

The Ir-Imes catalyst was activated by bubbling H$_2$ gas through the sample tube at a rate of 16.15 ccm for 1 hour at 40 °C. For each SABRE experiment, the sample was bubbled with 80-90% *p*-H$_2$ gas (produced with a Bruker BPHG 90 generator) at a rate of 43.46 ccm for 5s in the presence of the polarizing field. After this polarization build-up period, bubbling was shut off, and the electromagnet was triggered to output a field with a magnitude of ≈50 μT to control the transfer into the lab field. The sample was then transferred into a Magritek Spinsolve benchtop 1 T spectrometer and $^{15}$N and $^{1}$H spectra were acquired back-to-back. A repetition time of 200 s was allowed between SABRE experiments to allow for hyperpolarized $^{15}$N magnetization to fully relax.

A thermal reference sample containing 2.72 M $^{15}$N-pyridine in methanol-d$_4$ was created and a 250-scan-averaged reference spectrum was recorded using the same NMR tube and measurement magnet as the presented experiments. The signal intensity from the thermal reference was scaled down by a factor of 8.5 (assuming ½ of SABRE sample solvent evaporated during activation to create a final SABRE sample concentration of 320 mM) to control for concentration differences. Enhancement values were obtained by dividing SABRE sample signal by this concentration-scaled thermal value.

**B. Simulation methodology:**

The simulation framework used here to model SABRE dynamics has been reported in ref. [79]. The $^{15}$N-pyridine SABRE complex was modeled as a five-spin system which includes two $^{1}$H (*H*) from bound *p*-H$_2$ as well as one $^{15}$N (*S*) and two ortho $^{1}$H (*L*) from a bound pyridine. The SABRE complex was treated as a Y system, though it is known that pyridine creates an X geometry in which it binds both available equatorial positions of the SABRE catalyst. Simulations that accounted for the X geometry showed analogous dynamics to the Y geometry (just with twice the efficiency since pyridine is bound to both positions), so Y geometry was preferred for computational efficiency. Unless otherwise specified, this system used the following J-coupling values: $J_{HH} = -7$ Hz, $J_{HS} = 19.6$ Hz, $J_{HL} = 1.2$ Hz, $J_{LS} = -10.06$ Hz, $J_{LL} = -0.16$ Hz. Simulations included a 30.75 ppm chemical shift between *H* and *L*. The longitudinal relaxation times were assumed to be: $T_{1,H} = 1$ s, $T_{1,L} = 4$ s, $T_{1,S} = 40$ s. As stated in the main text, the simulations accounted for a ligand exchange rate of 16 s$^{-1}$, a hydride exchange rate of 2 s$^{-1}$, a 1:20 catalyst to ligand ratio, and assumed that *p*-H$_2$ flowed continuously ([Ir]/[H$_2$] is negligible). Unless otherwise specified, SABRE performance was evaluated for a given test condition by allowing the system to evolve in the polarizing field for 5 s before a $90_y$ pulse was applied to either $^{15}$N or $^{1}$H and a simulated FID was acquired by allowing a further 5 s of evolution to proceed at $B = 1$ T.

## C. Extended spin coherence pathways starting from $\vec{I}_1 \bullet \vec{I}_2 \longrightarrow L_z S_z$:

$$\vec{I}_1 \bullet \vec{I}_2 \xrightarrow{J_{HL}(I_{1y}L_y)} iI_{1z}I_{2x}L_y \xrightarrow{J_{HH}(I_{1x}I_{2x})} -I_{1y}L_y \xrightarrow{J_{LS}(L_zS_z)} -iI_{1y}L_xS_z \xrightarrow{J_{HL}(I_{1y}L_y)} L_zS_z \quad [S1]$$

$$\vec{I}_1 \bullet \vec{I}_2 \xrightarrow{J_{HL}(I_{1y}L_y)} iI_{1z}I_{2x}L_y \xrightarrow{J_{HH}(I_{1x}I_{2x})} -I_{1y}L_y \xrightarrow{J_{HS}(I_{1z}S_z)} -iI_{1x}L_yS_z \xrightarrow{J_{HL}(I_{1x}L_x)} -L_zS_z \quad [S2]$$

$$\vec{I}_1 \bullet \vec{I}_2 \xrightarrow{J_{HL}(I_{1y}L_y)} iI_{1z}I_{2x}L_y \xrightarrow{J_{LS}(L_zS_z)} -I_{1z}I_{2x}L_xS_z \xrightarrow{J_{HH}(I_{1x}I_{2x})} -iI_{1y}L_xS_z \xrightarrow{J_{HL}(I_{1y}L_y)} L_zS_z \quad [S3]$$

$$\vec{I}_1 \bullet \vec{I}_2 \xrightarrow{J_{HL}(I_{1y}L_y)} -iI_{1x}I_{2z}L_y \xrightarrow{J_{HH}(I_{1z}I_{2z})} -I_{1y}L_y \xrightarrow{J_{LS}(L_zS_z)} -iI_{1y}L_xS_z \xrightarrow{J_{HL}(I_{1y}L_y)} L_zS_z \quad [S4]$$

$$\vec{I}_1 \bullet \vec{I}_2 \xrightarrow{J_{HL}(I_{1y}L_y)} -iI_{1x}I_{2z}L_y \xrightarrow{J_{HH}(I_{1z}I_{2z})} -I_{1y}L_y \xrightarrow{J_{HS}(I_zS_z)} -iI_{1x}L_yS_z \xrightarrow{J_{HL}(I_{1x}L_x)} -L_zS_z \quad [S5]$$

$$\vec{I}_1 \bullet \vec{I}_2 \xrightarrow{J_{HL}(I_{1y}L_y)} -iI_{1x}I_{2z}L_y \xrightarrow{J_{LS}(L_zS_z)} I_{1x}I_{2z}L_xS_z \xrightarrow{J_{HH}(I_{1z}I_{2z})} -iI_{1y}L_xS_z \xrightarrow{J_{HL}(I_{1y}L_y)} L_zS_z \quad [S6]$$

$$\vec{I}_1 \bullet \vec{I}_2 \xrightarrow{J_{HL}(I_{1y}L_y)} -iI_{1x}I_{2z}L_y \xrightarrow{J_{HS}(I_{1z}S_z)} -I_{1y}I_{2z}L_yS_z \xrightarrow{J_{HH}(I_{1z}I_{2z})} -iI_{1x}L_yS_z \xrightarrow{J_{HL}(I_{1x}L_x)} -L_zS_z \quad [S7]$$

$$\vec{I}_1 \bullet \vec{I}_2 \xrightarrow{J_{HL}(I_{1x}L_x)} -iI_{1z}I_{2y}L_x \xrightarrow{J_{HH}(I_{1y}I_{2y})} -I_{1x}L_x \xrightarrow{J_{LS}(L_zS_z)} iI_{1x}L_yS_z \xrightarrow{J_{HL}(I_{1x}L_x)} L_zS_z \quad [S8]$$

$$\vec{I}_1 \bullet \vec{I}_2 \xrightarrow{J_{HL}(I_{1x}L_x)} -iI_{1z}I_{2y}L_x \xrightarrow{J_{HH}(I_{1y}I_{2y})} -I_{1x}L_x \xrightarrow{J_{HS}(I_zS_z)} iI_{1y}L_xS_z \xrightarrow{J_{HL}(I_{1y}L_y)} -L_zS_z \quad [S9]$$

$$\vec{I}_1 \bullet \vec{I}_2 \xrightarrow{J_{HL}(I_{1x}L_x)} -iI_{1z}I_{2y}L_x \xrightarrow{J_{LS}(L_zS_z)} -I_{1z}I_{2y}L_yS_z \xrightarrow{J_{HH}(I_{1y}I_{2y})} iI_{1x}L_yS_z \xrightarrow{J_{HL}(I_{1x}L_x)} L_zS_z \quad [S10]$$

$$\vec{I}_1 \bullet \vec{I}_2 \xrightarrow{J_{HL}(I_{1x}L_x)} iI_{1y}I_{2z}L_x \xrightarrow{J_{HH}(I_{1z}I_{2z})} -I_{1x}L_x \xrightarrow{J_{LS}(L_zS_z)} iI_{1x}L_yS_z \xrightarrow{J_{HL}(I_{1x}L_x)} L_zS_z \quad [S11]$$

$$\vec{I}_1 \bullet \vec{I}_2 \xrightarrow{J_{HL}(I_{1x}L_x)} iI_{1y}I_{2z}L_x \xrightarrow{J_{HH}(I_{1z}I_{2z})} -I_{1x}L_x \xrightarrow{J_{HS}(I_zS_z)} iI_{1y}L_xS_z \xrightarrow{J_{HL}(I_{1y}L_y)} -L_zS_z \quad [S12]$$

$$\vec{I}_1 \bullet \vec{I}_2 \xrightarrow{J_{HL}(I_{1x}L_x)} iI_{1y}I_{2z}L_x \xrightarrow{J_{LS}(L_zS_z)} I_{1y}I_{2z}L_yS_z \xrightarrow{J_{HH}(I_{1z}I_{2z})} iI_{1x}L_yS_z \xrightarrow{J_{HL}(I_{1x}L_x)} L_zS_z \quad [S13]$$

$$\vec{I}_1 \bullet \vec{I}_2 \xrightarrow{J_{HL}(I_{1x}L_x)} iI_{1y}I_{2z}L_x \xrightarrow{J_{HS}(I_{1z}S_z)} -I_{1x}I_{2z}L_xS_z \xrightarrow{J_{HH}(I_{1z}I_{2z})} iI_{1y}L_xS_z \xrightarrow{J_{HL}(I_{1y}L_y)} -L_zS_z \quad [S14]$$

$$\vec{I}_1 \bullet \vec{I}_2 \xrightarrow{J_{HL}(I_{1z}L_z)} I_{1x}I_{2y}L_z \longrightarrow \text{no pathways} \quad [S15]$$

$$\vec{I}_1 \bullet \vec{I}_2 \xrightarrow{J_{HS}(I_{1z}S_z)} iI_{1x}I_{2y}S_z \xrightarrow{J_{HH}(I_{y1}I_{y2})} -I_{1z}S_z \xrightarrow{J_{HL}(I_{1x}L_x)} -iI_{1y}L_xS_z \xrightarrow{J_{HL}(I_{1y}L_y)} L_zS_z \quad [S16]$$

$$\vec{I}_1 \bullet \vec{I}_2 \xrightarrow{J_{HS}(I_{1z}S_z)} -iI_{1y}I_{2x}S_z \xrightarrow{J_{HH}(I_{x1}I_{x2})} -I_{1z}S_z \xrightarrow{J_{HL}(I_{1x}L_x)} -iI_{1y}L_xS_z \xrightarrow{J_{HL}(I_{1y}L_y)} L_zS_z \quad [S17]$$

$$\vec{I}_1 \bullet \vec{I}_2 \xrightarrow{J_{HS}(I_{1z}S_z)} iI_{1x}I_{2y}S_z \xrightarrow{J_{HL}(I_{1y}L_y)} -I_{1z}I_{2y}S_zL_y \xrightarrow{J_{HH}(I_{1y}I_{2y})} iI_{1x}L_yS_z \xrightarrow{J_{HL}(I_{1x}L_x)} L_zS_z \quad [S18]$$

$$\vec{I}_1 \bullet \vec{I}_2 \xrightarrow{J_{HS}(I_{1z}S_z)} -iI_{1y}I_{2x}S_z \xrightarrow{J_{HL}(I_{1x}L_x)} -I_{1z}I_{2x}S_zL_x \xrightarrow{J_{HH}(I_{1x}I_{2x})} -iI_{1y}L_xS_z \xrightarrow{J_{HL}(I_{1y}L_y)} L_zS_z \quad [S19]$$

$$\vec{I}_1 \bullet \vec{I}_2 \xrightarrow{J_{HS}(I_{1z}S_z)} iI_{1x}I_{2y}S_z \xrightarrow{J_{HH}(I_{y1}I_{y2})} -I_{1z}S_z \xrightarrow{J_{HL}(I_{1y}L_y)} iI_{1x}L_yS_z \xrightarrow{J_{HL}(I_{1x}L_x)} L_zS_z \quad [S20]$$

$$\vec{I}_1 \bullet \vec{I}_2 \xrightarrow{J_{HS}(I_{1z}S_z)} -iI_{1y}I_{2x}S_z \xrightarrow{J_{HH}(I_{x1}I_{x2})} -I_{1z}S_z \xrightarrow{J_{HL}(I_{1y}L_y)} iI_{1x}L_yS_z \xrightarrow{J_{HL}(I_{1x}L_x)} L_zS_z \quad [S21]$$

## Fig. S1. Experimental $^1$H data from $^{15}$N-pyridine SABRE sample

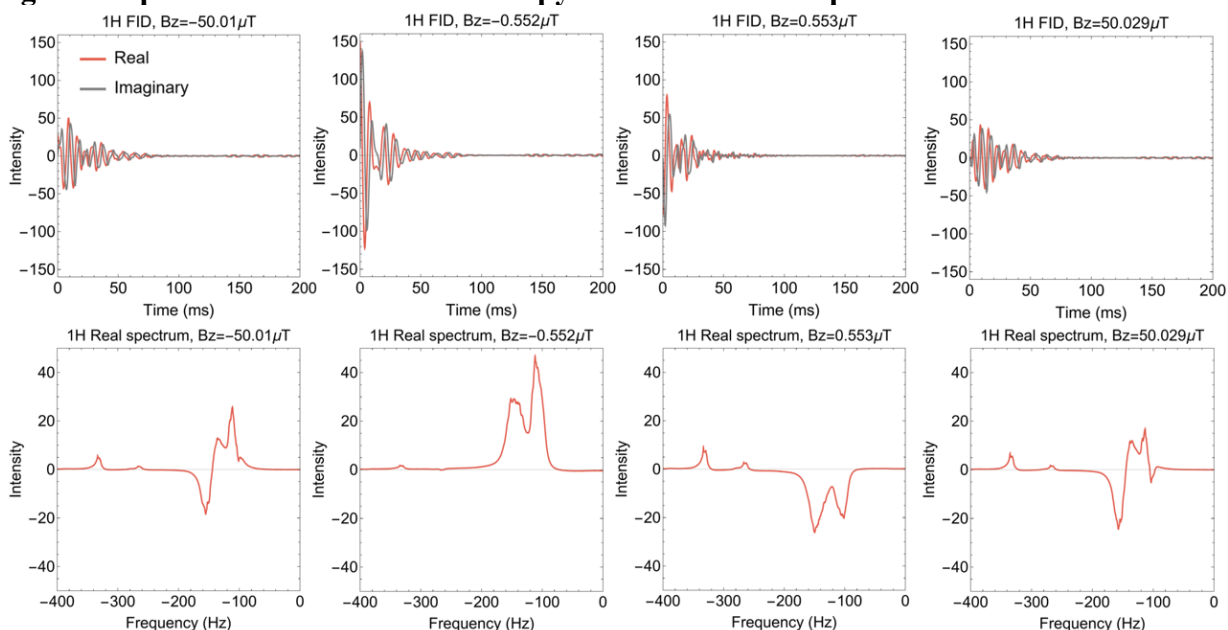

These $^1$H spectra were taken immediately after the $^{15}$N spectra shown in main text Fig. 3. The sample composition and experimental procedure for this data set are detailed in supplemental section A above. While the presence of the H$_2$ bubbler in the NMR tube introduces significant inhomogeneity and reduces resolution, the two peaks at $-340$ Hz and $-260$ Hz may be assigned to methanol, while the group centered at $-140$ Hz may be assigned to the ligand pyridine protons. Similar to what is observed in the $^{15}$N signal, $^1$H signal far from the SHEATH resonance is antiphase and does not invert when the sign of the field changes. $^1$H signal at the SHEATH resonance is predominantly one-spin magnetization which does invert when the sign of the field changes. While the SHEATH condition is far from the $^1$H-SABRE resonance (which is maximized at 6.5 mT), polarization may still be transferred to the ligand $^1$H from the matched $^{15}$N through the $J_{LS}$ coupling during polarization build-up to generate observable $^1$H magnetization.

## Fig. S2. Magnitude of $^{15}$N-pyridine out-of-phase magnetization, simulation and experiment

This plot compares the magnitude of out-of-phase signal (signal at $t = 1/4J_{LS}$) between experimental $^{15}$N-pyridine (purple, absolute value of the data shown in main text Fig. 3, left) and simulation (gray, same parameters detailed in SI section B). Numerical simulations confirm the presence of strong effective magnetization far from resonance and predict that the magnitude of out-of-phase magnetization will be 1/5$^{th}$ the intensity of in-phase magnetization, which matches experimental observations.

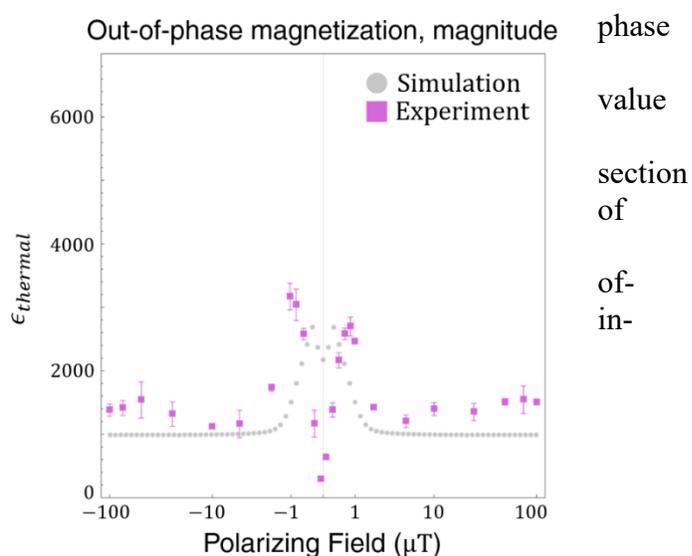

**Fig. S3. Effects of supplying $\vec{I}_1 \cdot \vec{I}_2$ vs $I_{1z}I_{2z}$ order in $^1$H-SABRE ($L$ only)**

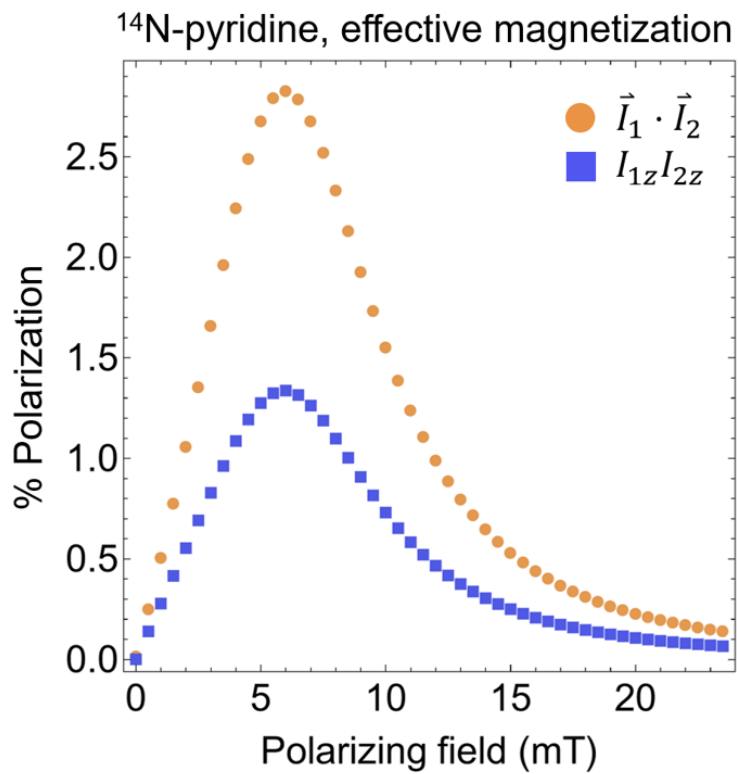

These simulations were run according to the methodology outlined in supplemental section B, except that the system omitted the $^{15}$N spin on pyridine and only accounted for couplings to two ortho-$^1$H (resulting in a four-spin instead of a five-spin system). The plot records the effective magnetization generated at a certain polarizing field, where effective magnetization is defined as the maximum value of the magnitude of the $^1$H FID which was acquired after 5 s of polarization build-up. In the orange curve, the $p$-H$_2$ density matrix was defined as $\frac{1}{4}\hat{1} - (\vec{I}_1 \cdot \vec{I}_2)$, which represents pure singlet order. In the blue curve, $p$-H$_2$ is modeled as $\frac{1}{4}\hat{1} - I_{1z}I_{2z}$, representing a mixture of singlet and triplet states. We observe that in the $^1$H SABRE regime, for a ligand which does not contain a strongly-coupled heteroatom, hyperpolarization efficiency decreases by about a factor of 2 in the absence of the $I_{1x}I_{2x}$ and $I_{1y}I_{2y}$ terms.

**Fig. S4. Effects of supplying $\vec{I}_1 \cdot \vec{I}_2$ vs $I_{1z}I_{2z}$ order in SABRE-SHEATH ($L$ and $S$)**

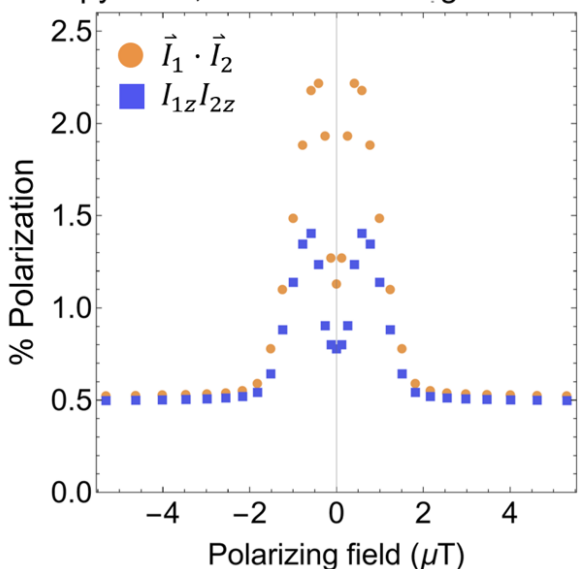 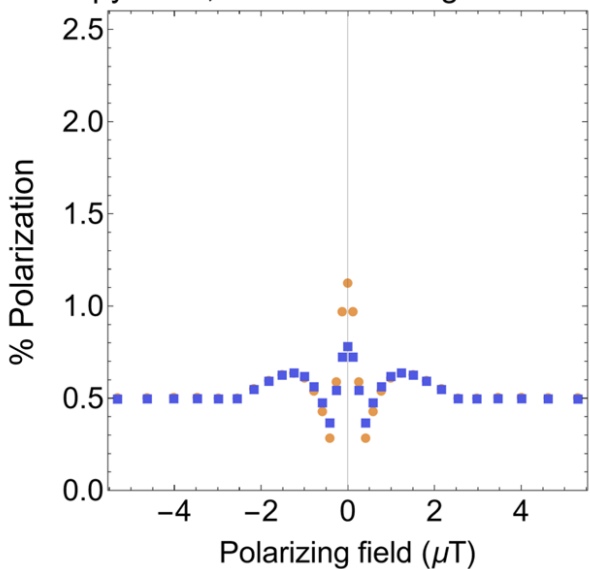

This figure shows analogous data to Fig. S3, but for the $^{15}$N-pyridine system which contains couplings to both $L$ and $S$ (see SI section B for simulation parameters), where the polarizing field sweep is now centered on the X-SABRE resonance. Similar to what is observed in main text Fig. 7, the effective magnetization resulting from two-spin order is robust to the interconversion of singlet/triplet order on the hydride. Production of one-spin order, which dominates at the SHEATH resonance (≈ 0.5 µT), is hindered when the $I_{1x}I_{2x}$ and $I_{1y}I_{2y}$ terms are removed, but to a lesser extent than what is observed in the $^{14}$N-pyridine $^1$H-SABRE case (Fig. S3).

### Fig. S5. Nonresonant SABRE demonstrated using $^{15}$N-acetonitrile

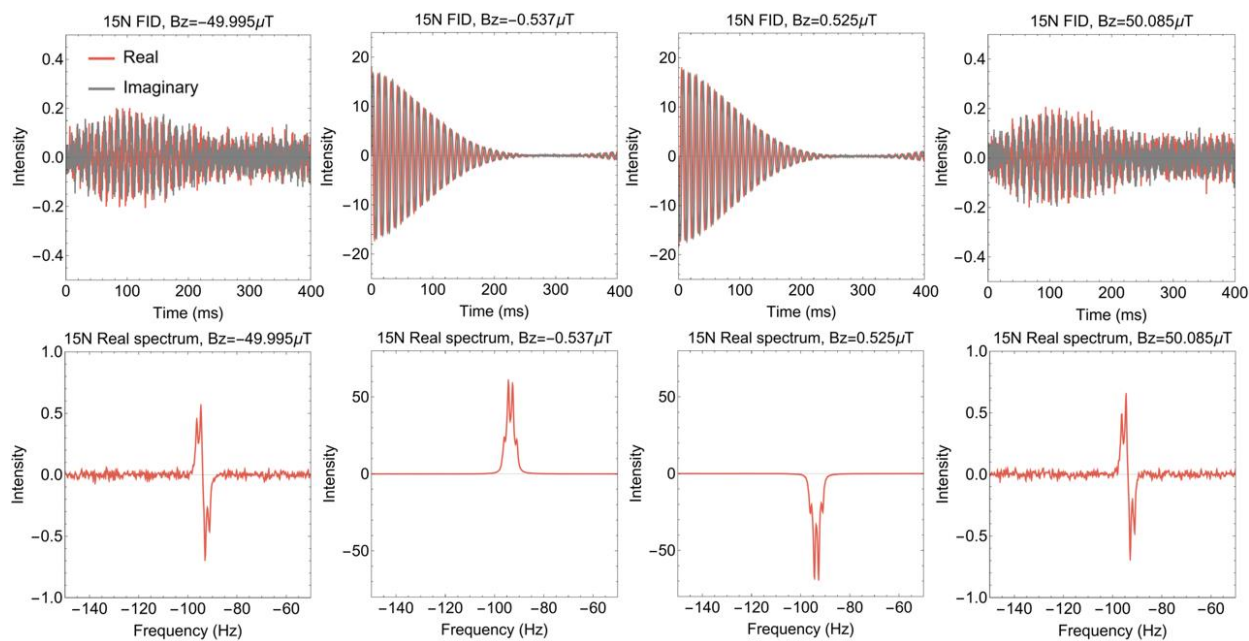

A $^{15}$N-acetonitrile SABRE sample was prepared with 150 mM $^{15}$N-acetonitrile (Cambridge Isotopes Laboratories NLM-175-0.5, CAS 14149-39-4), 50 mM $^{14}$N-pyridine as a spin-silent co-ligand (provided by collaborator, CAS 110-86-1), and 6.5 mM Ir-Imes in 250 µL methanol-d$_4$. Sample was activated by bubbling H$_2$ gas through the sample tube at a rate of 16.15 ccm for 30 min at 25 °C. Data acquisition followed the same procedure as described in supplemental section A, with the exception that polarization build-up was allowed to occur for 30 s, as opposed to 5 s. As was observed in the $^{15}$N-pyridine experiments, the two-spin magnetization did not invert when the polarizing field was inverted, while the one-spin magnetization did. In acetonitrile, the signal intensity resulting from two-spin order was observed to be about 100 times less intense than the one-spin order. This difference is significantly greater than the roughly ten-fold difference that was observed in pyridine (Fig. 3.). Since the two-spin order has a shorter lifetime than the one-spin order (noted in main text), this difference in signal is likely a result of the longer, 30 s polarization time, as the two-spin order would reach a plateau in polarization build-up faster than the one-spin order.